\documentclass[iop]{emulateapj}

\usepackage{natbib} 
\usepackage{amsmath}
\usepackage{color}
\usepackage{booktabs}
\usepackage{multirow}
\usepackage{chngcntr}
\usepackage{subfig}

\definecolor{orange}{rgb}{0.93, 0.57, 0.13}
\definecolor{green}{rgb}{0.01, 0.75, 0.24}

\newcommand{\ie}{\emph{i.e.}}
\newcommand{\eg}{e.g.}
\newcommand{\emm}[1]{\ensuremath{#1}}   
\newcommand{\emr}[1]{\emm{\mathrm{#1}}} 

\newcommand{\Tdust}{\emm{T_\emr{dust}}}
\newcommand{\Tgas}{\emm{T_\emr{gas}}}
\newcommand{\Tex}{\emm{T_\emr{ex}}}

\newcommand{\Rout}{\emm{R_\emr{out}}}

\newcommand{\Msun}{\emr{M_\odot}}
\newcommand{\Mstar}{\emr{M_\star}}
\newcommand{\Lsun}{\emr{L_\odot}}
\newcommand{\Lstar}{\emr{L_\star}}
\newcommand{\Mrate}{\emr{\dot{M_\star}}}
\newcommand{\Mdisk}{\emr{M_\emr{disk}}}

\newcommand{\thco}{\emr{^{13}CO}}                  
\newcommand{\ceio}{\emr{C^{18}O}}                  
\newcommand{\hthcn}{\emr{H^{13}CN}}                  
\newcommand{\hcfin}{\emr{HC^{15}N}}                  

\newcommand{\Nratio}{\emr{^{14}N/^{15}N}}                   %
\newcommand{\Cratio}{\emr{^{12}C/^{13}C}}                   %

\renewcommand{\deg}{\emm{^\circ}}
\newcommand{\pccm}{~\rm{cm}^{-3}}

\newcommand{\ps}{~\rm{s}^{-1}}
\newcommand{\kms}{\emr{\,km\,s^{-1}}}

\begin{document}


\title{Nitrogen Fractionation in protoplanetary disks from the \hthcn/\hcfin{} ratio}


\author{V.V. Guzm\'an\altaffilmark{$\star$}}
\email{vguzman@cfa.harvard.edu}

\author{K.I. \"Oberg}

\author{J. Huang}

\author{R. Loomis}

\author{C. Qi}

\affil{Harvard-Smithsonian Center for Astrophysics, 60 Garden Street, Cambridge, MA 02138, USA}

\altaffiltext{$\star$}{Currently at Joint Atacama Large Millimeter/submillimeter Array (ALMA) Observatory, Avenida Alonso de C\'ordova 3107, Vitacura, Santiago, Chile.}

\begin{abstract}
Nitrogen fractionation is commonly used to assess the thermal history
of Solar System volatiles. With ALMA it is for the first time possible
to directly measure \Nratio{} ratios in common molecules during
assembly of planetary systems. We present ALMA observations of the
\hthcn{} and \hcfin{} $J=3-2$ lines at $0''.5$ angular resolution,
toward a sample of six protoplanetary disks, selected to span a range
of stellar and disk structure properties. Adopting a typical \Cratio{}
ratio of 70, we find comet-like \Nratio{} ratios of $80-160$ in 5/6 of
the disks (3 T Tauri and 2 Herbig Ae disks) and lack constraints for
one of the T Tauri disks (IM Lup). There are no systematic differences
between T Tauri and Herbig Ae disks, or between full and transition
disks within the sample. In addition, no correlation is observed
between disk-averaged D/H and \Nratio{} ratios in the sample. One of
the disks, V4046 Sgr, presents unusually bright HCN isotopologue
emission, enabling us to model the radial profiles of \hthcn{} and
\hcfin{}. We find tentative evidence of an increasing \Nratio{} ratio
with radius, indicating that selective photodissociation in the inner
disk is important in setting the \Nratio{} ratio during planet
formation.

\end{abstract}






\newcommand{\TabSources}{%
  \begin{table*}
    \begin{center}
    \caption{Star and disk properties in the sample.} 
    \label{tab:sample}
        \begin{tabular}{lccclcccccc}\toprule
          Source & Distance & Spectral  & Age & \Mstar & \Lstar & \Mrate & Disk Incl. & Disk PA & \Mdisk & $v_{LSR}$  \\
                 & pc       & type      & Myr & \Msun  & \Lsun  & $10^{-9}$ \Msun yr$^{-1}$ & deg & deg & \Msun & km s$^{-1}$\\
          \midrule
          \multicolumn{11}{c}{\it T~Tauri}\\
          AS~209    & 126 & K5    & 1.6 & 0.9 & 1.5     & 51   & 38 & 86 & 0.015 & 4.6 \\
          IM~Lup    & 161 & M0    & 1   & 1.0 & 0.93    & 0.01 & 50 & 144.5 & 0.17 & 4.4 \\
          LkCa~15   & 140 & K3    & 3-5 & 0.97 & 0.74     & 1.3  & 52 & 60 & 0.05-0.1 & 6.3 \\
          V4046~Sgr &  72 & K5,K7 & 24  & 1.75 & 0.49,0.33 & 0.5  & 33.5 & 76 & 0.028 & 2.9 \\
          \midrule
          \multicolumn{11}{c}{\it Herbig Ae}\\
          MWC~480   & 142 & A4 & 7 & 1.65 & 11.5 & 126 & 37 & 148 & 0.11 & 5.1 \\
          HD~163296 & 122 & A1 & 4 & 2.25 & 30   & 69 & 48.5 & 132 & 0.17 & 5.8 \\
          \bottomrule
        \end{tabular}
    \end{center}
    Note: Table reproduced from \citet{huang2017}, where a
    complete list of references is given.
\end{table*}
}

\newcommand{\TabObs}{%
  \begin{table*}[t!]
    \begin{center}
      \caption{Line observations.}
    \label{tab:obs}
        \begin{tabular}{lcrcc|crcc}\toprule
          & \multicolumn{4}{c}{\hthcn{} $J=3-2$} & \multicolumn{4}{c}{\hcfin{} $J=3-2$} \\
          \cmidrule{2-9}
          \multirow{2}{*}{Source} & Beam & PA & Channel rms$^a$ & Mom. zero rms$^a$ & Beam & PA & Channel rms$^a$ & Mom. zero rms \\
          & arcsec & $^{\deg}$ & mJy beam$^{-1}$ & mJy~km~s$^{-1}$ & arcsec & $^{\deg}$ & mJy beam$^{-1}$ & mJy~km~s$^{-1}$\\ 
          \midrule          
          AS~209    & $0.51\times0.50$ & $-86.9$ & 4.3 &  7.6 & $0.51\times0.44$ & $-62.4$ & 3.4 & 6.5 \\
          IM~Lup    & $0.46\times0.43$ & $71.1$  & 3.4 &  5.6 & $0.47\times0.43$ & $71.8$  & 3.0 & 5.2 \\
          LkCa~15   & $0.65\times0.47$ & $-13.0$ & 3.2 &  6.3 & $0.67\times0.50$ & $-14.7$ & 2.9 & 5.7 \\
          V4046~Sgr & $0.60\times0.48$ & $87.3$  & 3.4 & 10.4 & $0.59\times0.48$ & $87.0$  & 3.2 & 9.4 \\
          MWC~480   & $0.74\times0.46$ & $-8.7$  & 3.3 &  7.0 & $0.74\times0.46$ & $-9.0$  & 3.1 & 6.7 \\
          HD~163296 & $0.57\times0.44$ & $-87.9$ & 2.6 &  6.9 & $0.59\times0.46$ & $-87.3$ & 2.4 & 6.4 \\
          \bottomrule
        \end{tabular}
    \end{center}
    $^{a}$ Channel rms noise in 0.5~km~s$^{-1}$ channel width.
\end{table*}
}

\newcommand{\TabSpecParam}{%
  \begin{table}[t!]
    \begin{center}
    \caption{Spectroscopic parameters.} 
    \label{tab:spec}
        \begin{tabular}{ccccc}\toprule
          Line & Frequency & $E_u/k$ & $A_{ul}$ & $g_u$ \\
          & GHz & K & $\ps$ \\
          \midrule
          \hthcn{} $J=3-2$ & 259.0118 & 24.86 & $7.7\times10^{-4}$ & 21 \\
          \hcfin{} $J=3-2$ & 258.1571 & 24.78 & $7.6\times10^{-4}$ & 7\\
          \bottomrule
        \end{tabular}
    \end{center}
  \end{table}
}

\newcommand{\TabParam}{%
  \begin{table}[b!]
    \begin{center}
    \caption{Adopted parameters in disk model.} 
    \label{tab:param}
        \begin{tabular}{lll}\toprule
          \multirow{2}{*}{Scale height} & $H_{10}$ & 0.4~AU \\
          & $h$ & 1.25\\
          \midrule
          \multirow{5}{*}{Dust surface density} &          $\Sigma_c$ & 0.226 g~cm$^{-2}$  \\
          & $\gamma$ & 1\\
          & $r_c$ & 75~AU\\
          & $\Sigma_{cav}$ & $10^{-4}$\\
          & $r_{cav}$ & 3~AU\\
          \midrule
          \multirow{3}{*}{Gas temperature} &          $T_{10}$ & 200~K \\
          & $q_{atm}$ & 0.8\\
          & $\delta$ & 2 \\
          \bottomrule
        \end{tabular}
    \end{center}
\end{table}
}

\newcommand{\TabFluxes}{%
  \begin{table*}[t!]
    \begin{center}
    \caption{Integrated fluxes and inferred abundance ratios in the disk sample.}
    \label{tab:ratios}
        \begin{tabular}{lrrcr}\toprule
          Source & $F(\hthcn)^a$ & $F(\hcfin)^a$ & $N(\hthcn)/N(\hcfin)^b$ & \Nratio{}$^c$ \\
          \midrule
          AS~209    & $ 175\pm25$ & $ 79\pm23$ & $2.2\pm0.7$ & $156\pm71$ \\ 
          IM~Lup    & $  <50^d  $ & $   <50^d$   & -           & -\\
          LkCa~15   & $ 120\pm20$ & $101\pm18$ & $1.2\pm0.3$ & $83 \pm32$  \\
          V4046~Sgr & $1098\pm35$ & $670\pm26$ & $1.6\pm0.1$ & $115\pm35$  \\
          MWC~480   & $ 132\pm13$ & $ 75\pm13$ & $1.8\pm0.3$ & $123\pm45$  \\
          HD~163296 & $ 177\pm23$ & $ 87\pm21$ & $2.0\pm0.6$ & $142\pm59$  \\
          \bottomrule
        \end{tabular}
    \end{center}
    $^{a}$ The uncertainties do not include absolute flux calibration
    errors.  $^{b}$ Computed assuming LTE and $\Tex=15$~K. $^{c}$
    Adopting $\Cratio=70$ and an associated uncertainty of
    30\%. $^{d}$ $3\sigma$ upper limits.
\end{table*}
}

\newcommand{\FigCOmodels}{%
\begin{figure}[t!]
  \centering
  \includegraphics[width=0.47\textwidth]{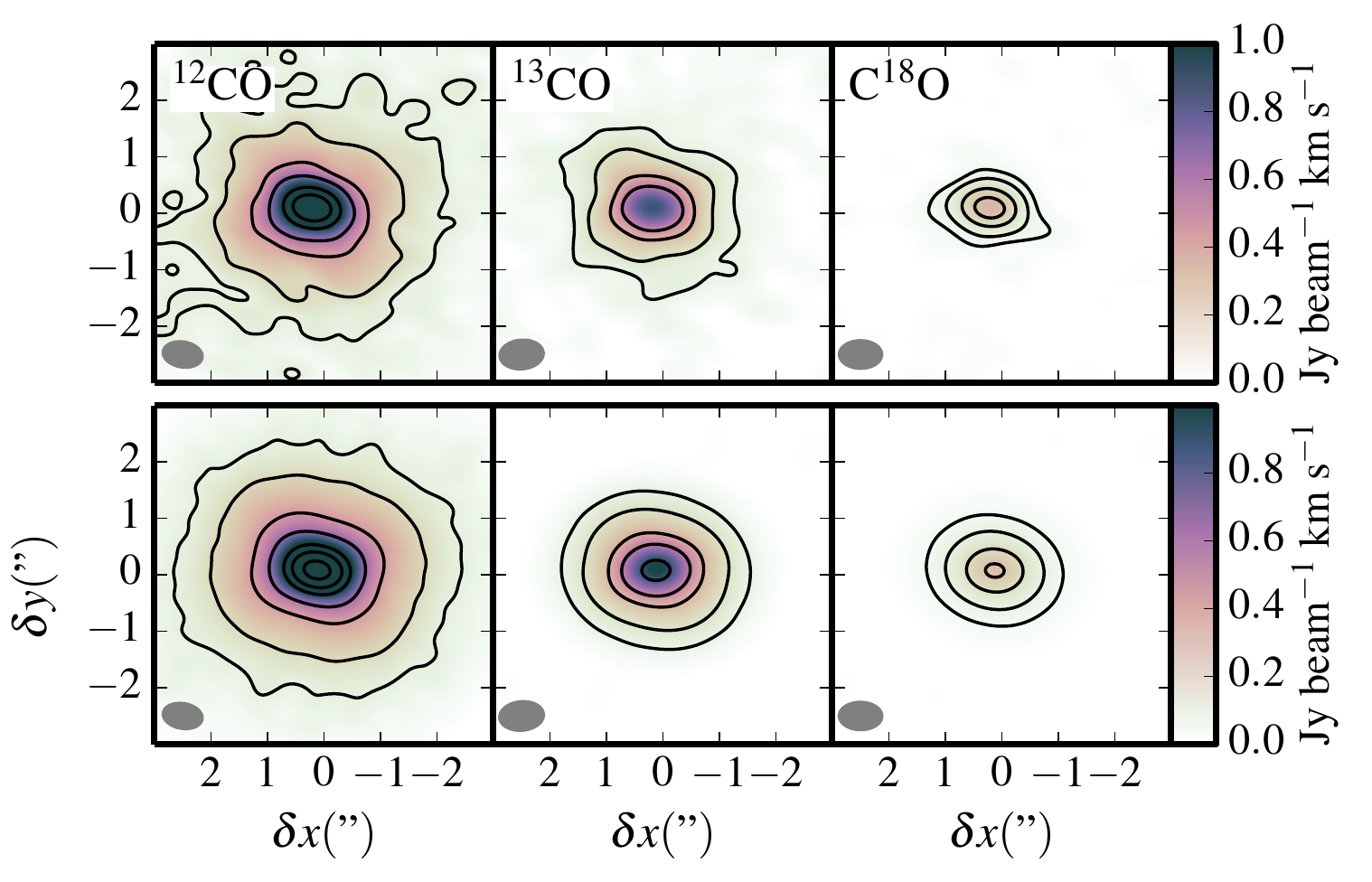}
  \caption{Moment zero maps of the CO, \thco{} and \ceio{} $J=2-1$
    lines in V4046~Sgr. The observed and modeled integrated emission
    maps are shown in the upper and lower panels, respectively. The
    contours levels correspond to 5, 10, 20, 30, 50 and
    100$\sigma$, where the rms is 22, 18 and
    10~mJy~beam$^{-1}$~\kms{} for CO, \thco{} and \ceio{},
    respectively.}
  \label{fig:COmodel}
\end{figure}
}

\newcommand{\FigSurvey}{%
\begin{figure*}[t!]
  \centering
  \includegraphics[trim=0 50 0 0,width=0.98\textwidth]{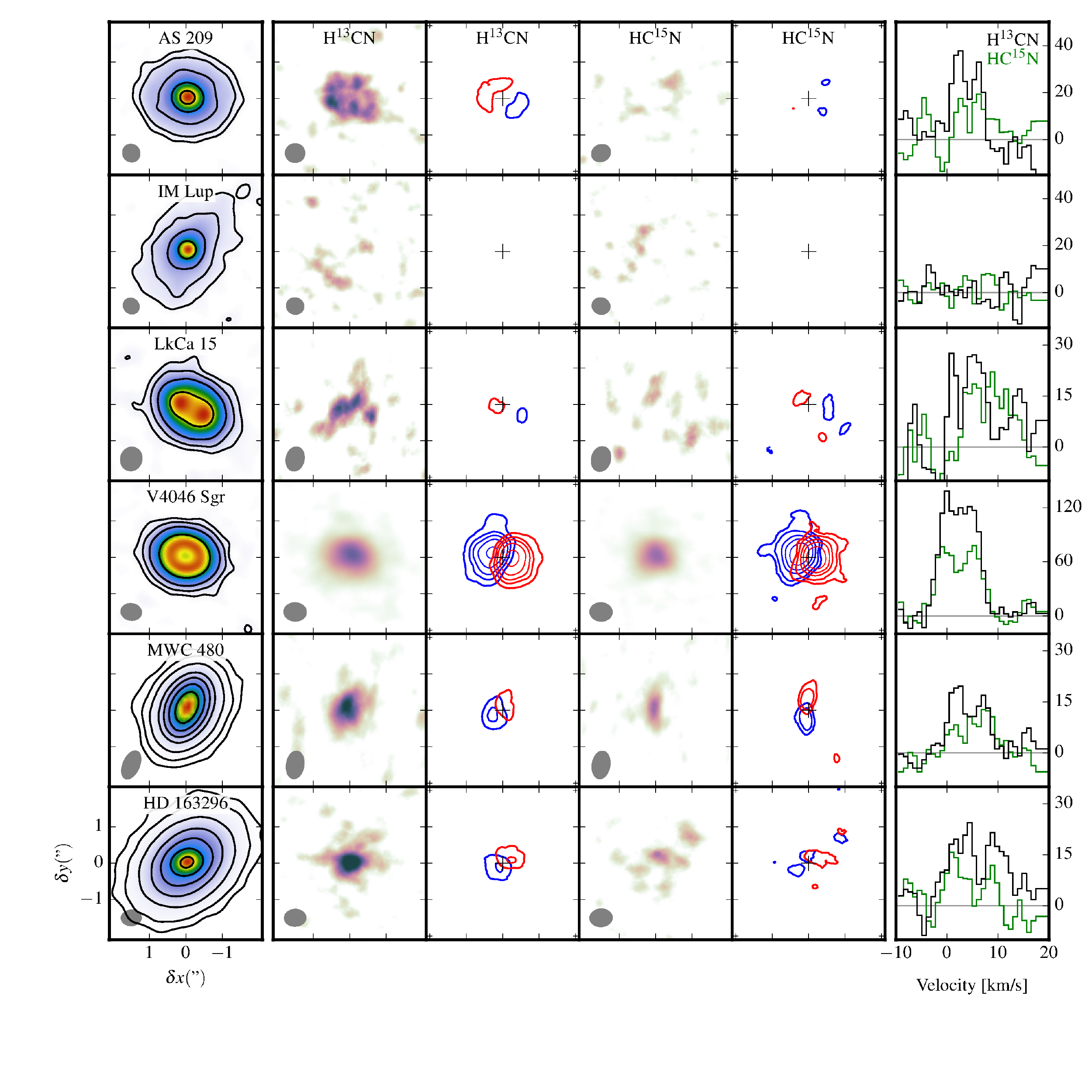}
  \caption{{\it Left:} Dust continuum emission. Contour levels
    correspond to 3, 10, 30, 100, 200, 400 and 800$\sigma$. {\it
      Middle:} Moment zero maps for \hthcn{} and \hcfin{}, integrated
    over the full line width (colors) and over the red- and
    blue-shifted parts of the line (contours). Noise below
    2~mJy~beam$^{-1}$ has been clipped to produce the moment zero
    maps. Color scales start at $2\sigma$. Contour levels correspond
    to 3, 5, 7, 10, 15, 20 and 30$\sigma$. For \hcfin{} contour levels
    start at $2\sigma$. {\it Right:} Disk integrated spectra,
    extracted using elliptical masks (see Appendix).}
  \label{fig:survey}
\end{figure*}
}

\newcommand{\FigResiduals}{%
\begin{figure*}[t!]
  \centering
  \includegraphics[width=\textwidth]{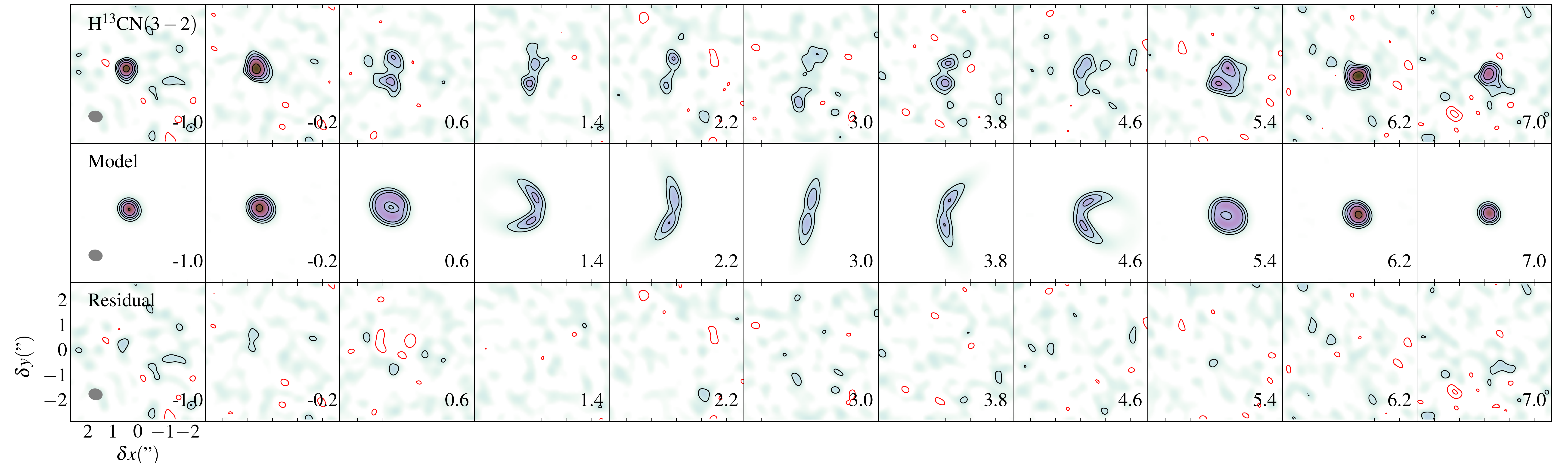}
  \includegraphics[width=\textwidth]{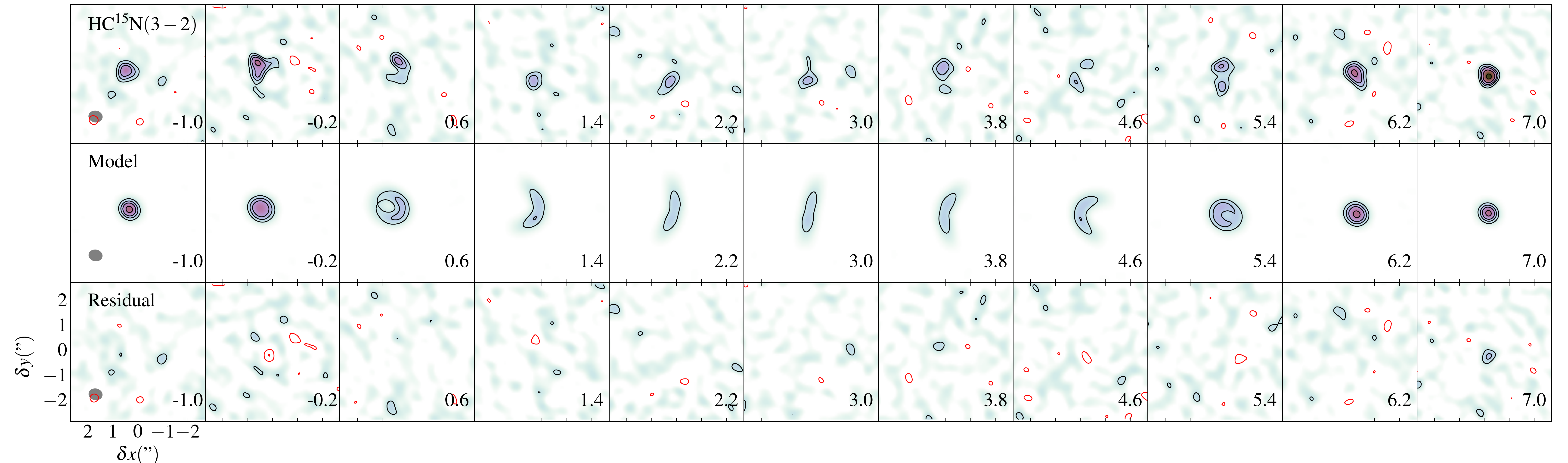}
  \caption{Channel maps of the observed \hthcn{} and \hcfin{} $J=3-2$
    emission lines in V4046~Sgr (upper panels). The middle panels show
    the best-fit models for each line. The residuals are shown in the
    bottom panels. Positive (black) and negative (red) contour
      levels correspond to 3, 5, 7 10, 15, 20 and 25$\sigma$.}
  \label{fig:residuals}
\end{figure*}
}

\newcommand{\FigIsotopologuesRatios}{%
\begin{figure}[b!]
  \centering
  \includegraphics[width=0.48\textwidth]{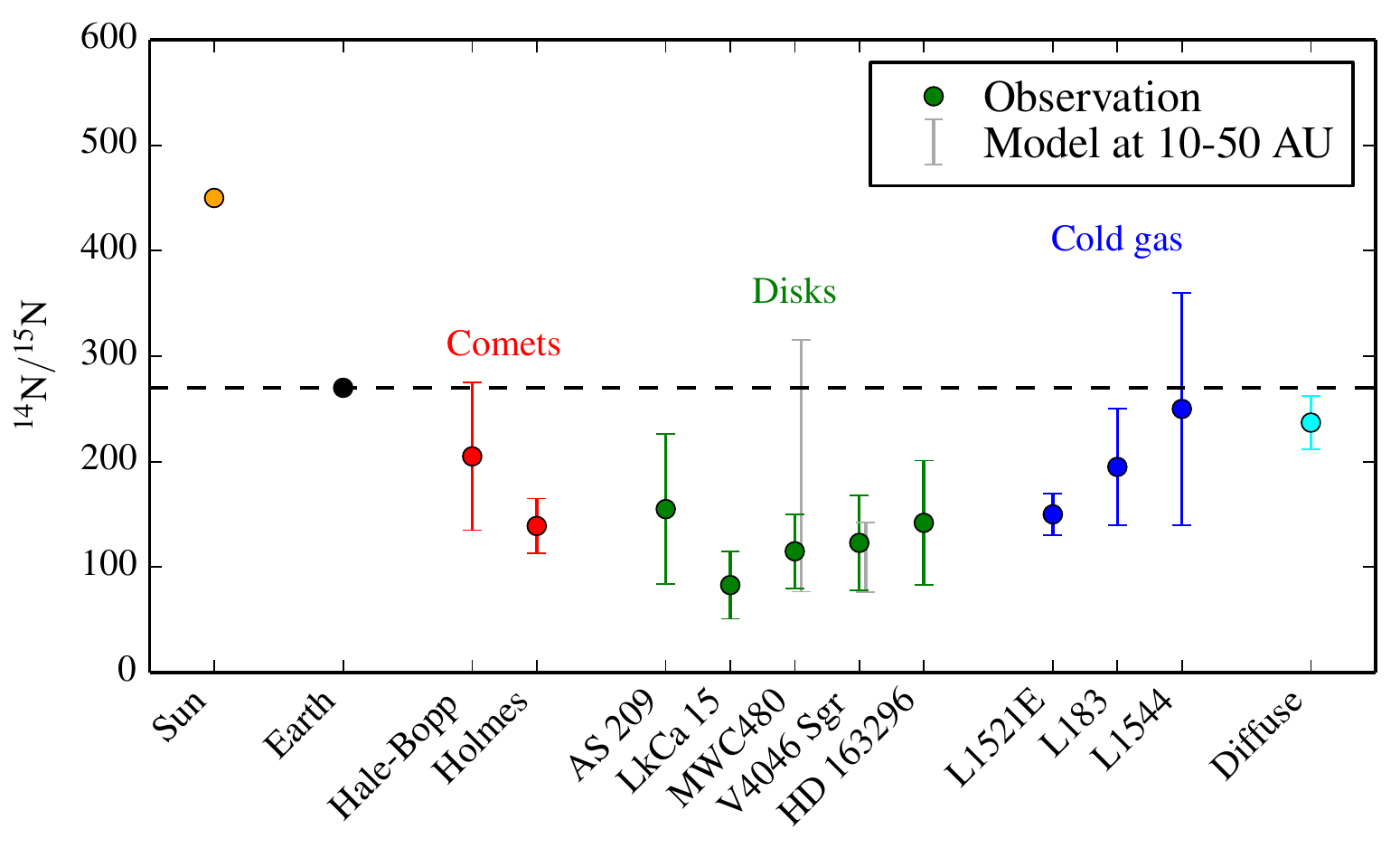}
  \caption{\Nratio{} isotopic ratios observed toward different solar
    system bodies \citep{marty2011,bockelee2008}, the ISM
    \citep{hily-blant2013,ikeda2002,lucas1998}, and protoplanetary
    disks (this work). The Solar and Earth values were measured for
    N$^+$ and N$_2$, respectively, the rest of the sources all
    correspond to isotopic ratios measured for HCN.}
  \label{fig:Nratios}
\end{figure}
}

\newcommand{\FigDNratio}{%
\begin{figure}[t!]
  \centering
  \includegraphics[width=0.48\textwidth]{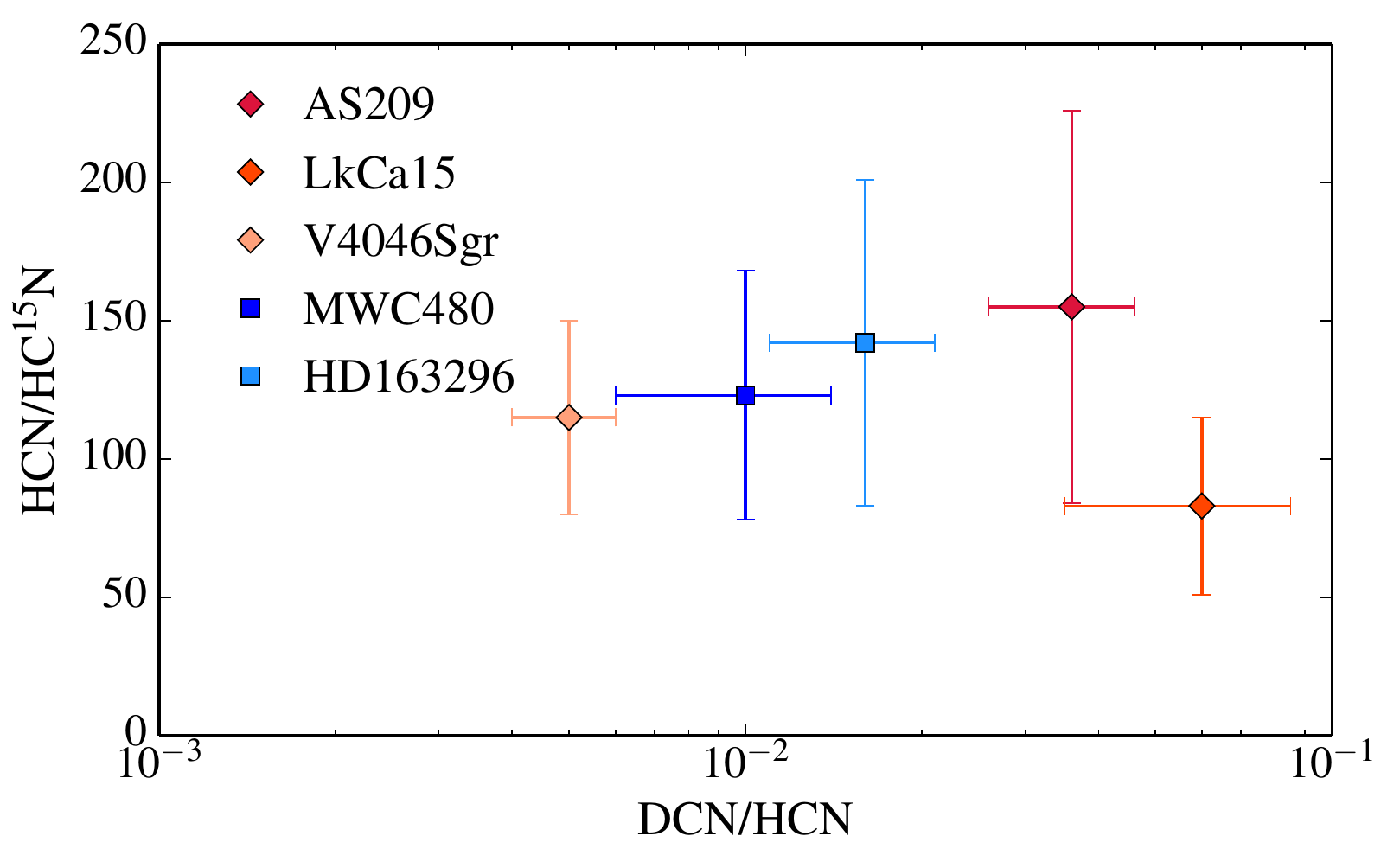}
  \caption{N fractionation as a function of deuterium
    fractionation. Disks around T~Tauri stars are shown in red colors,
    while disks around Herbig~Ae stars are shown in blue colors. The
    DCN/HCN ratios are taken from \citet{huang2017}. Both
    abundance ratios are computed assuming LTE conditions and
    $\Tex=15$~K.}
  \label{fig:DNratios}
\end{figure}
}

\newcommand{\FigDiskStructure}{%
\begin{figure}[t!]
  \centering
  \includegraphics[width=0.5\textwidth]{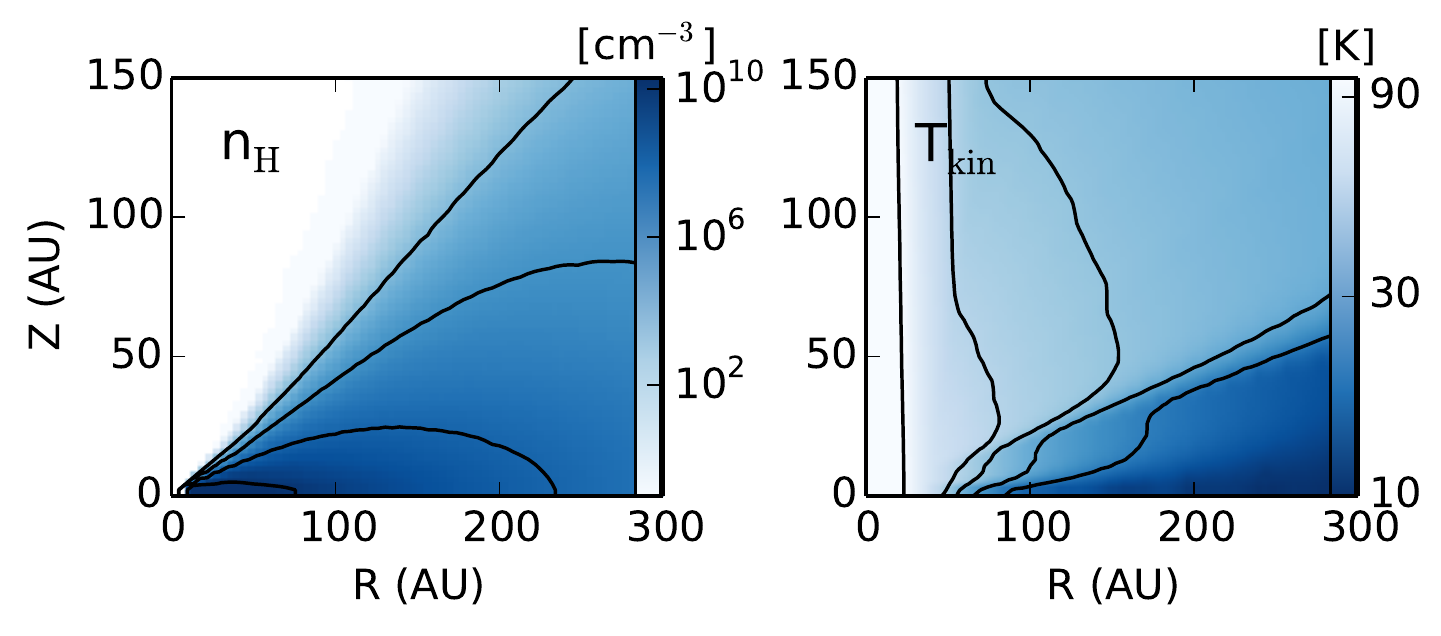}
  \caption{Disk structure for V4046~Sgr. The contours mark densities
    of $10^4$, $10^6$, $10^8$ and $10^{10} \pccm$ (left panel) and gas
    temperatures of 20, 30, 40, 50, 100~K (right panel).}
  \label{fig:structure}
\end{figure}
}

\newcommand{\FigTriangles}{%
\begin{figure*}[b!]
  \centering
  \includegraphics[width=0.44\textwidth]{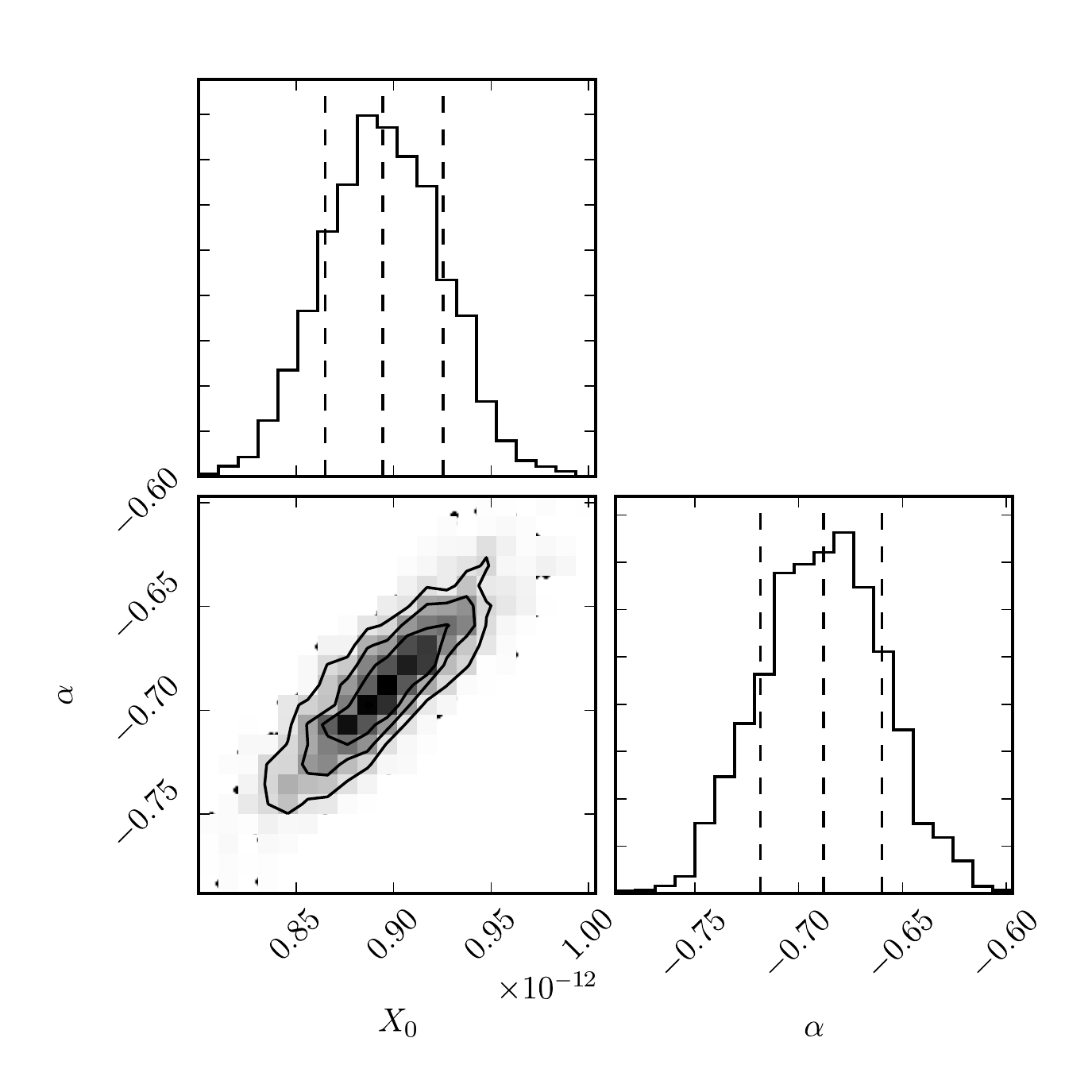}
  \includegraphics[width=0.44\textwidth]{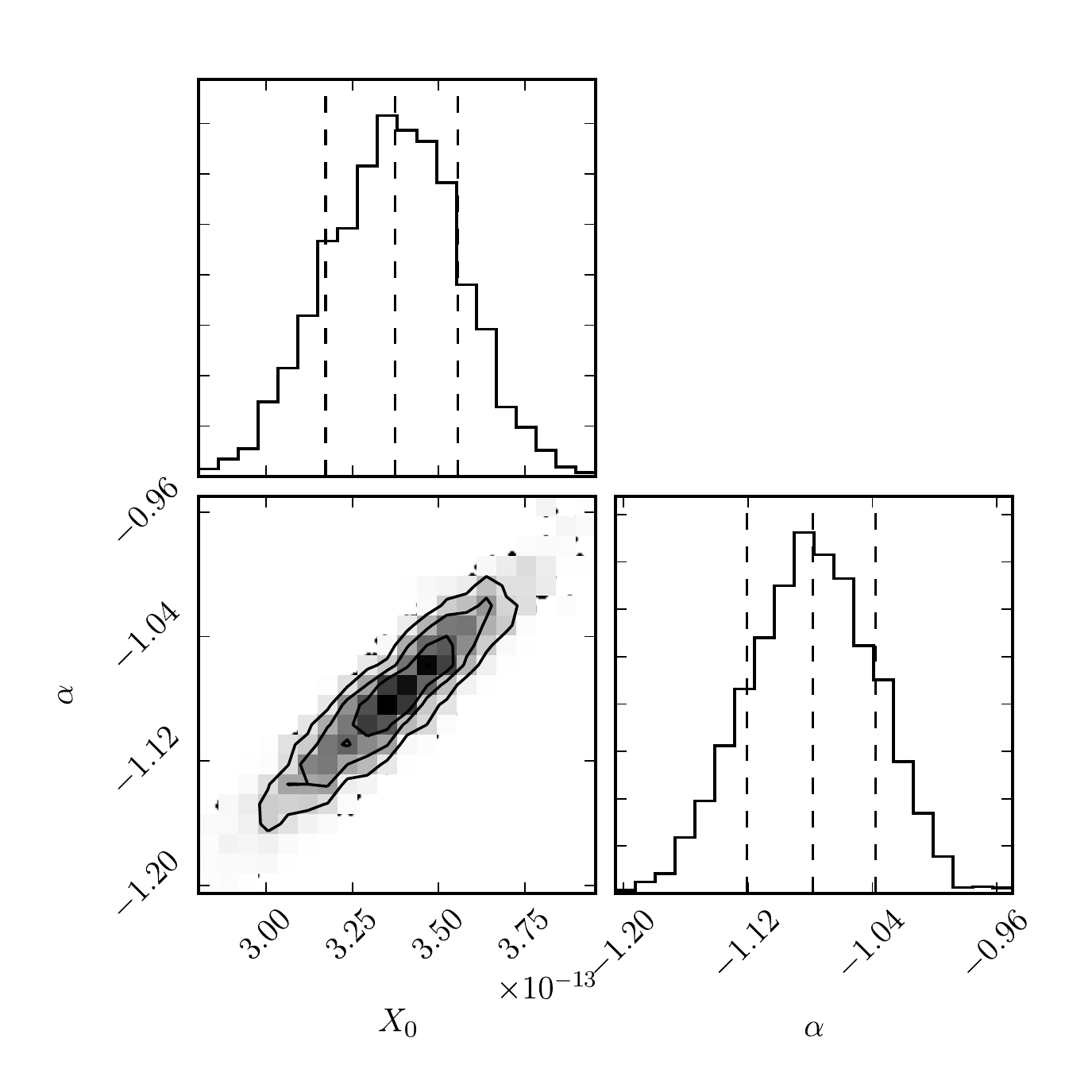}
  \caption{Line modeling results of \hthcn{} (left) are \hcfin{}
    (right) for the disk around V4046~Sgr. The gray color scales show
    the joint probability distribution probabilities for the
    parameters, $\alpha$ and $X_0$. The contours represent the
    1$\sigma$, 2$\sigma$ and 3$\sigma$ levels. The histograms show the
    marginal distribution for each parameter and the vertical dashed lines mark
    the median and 1$\sigma$ uncertainty.}
  \label{fig:mcmc}
\end{figure*}
}

\newcommand{\FigRadialProfile}{%
\begin{figure}[b!]
  \centering
  \includegraphics[width=0.5\textwidth]{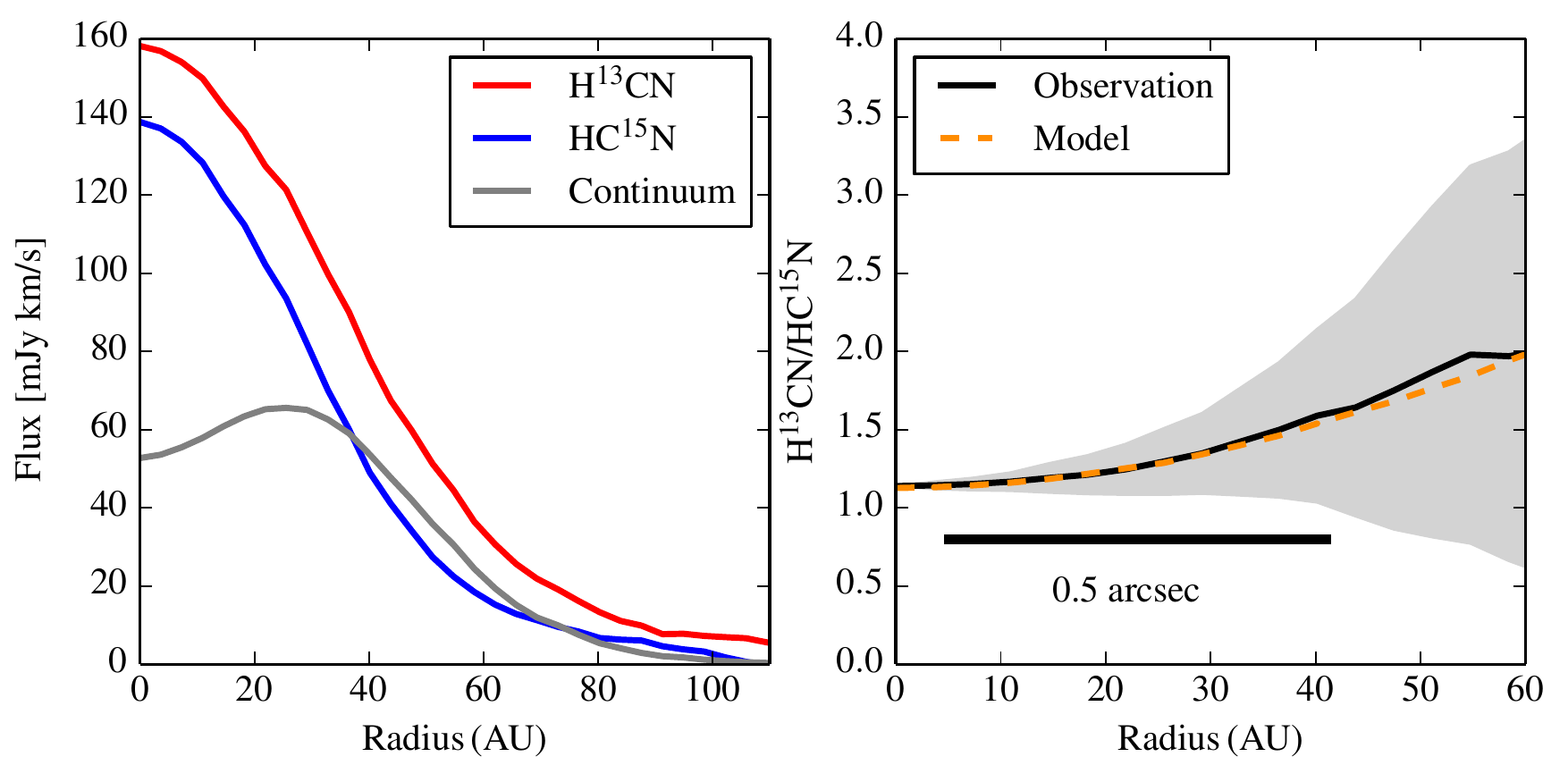}
  \caption{{\it Left:} Azimuthally-averaged emission profiles of the
    dust continuum emission (gray), \hthcn{} $J=3-2$ (red) and \hcfin{}
    $J=3-2$ (blue) lines in V4046~Sgr. {\it Right:} The observed
    \hthcn/\hcfin{} flux ratio (black) and the ratio derived from our
    best-fit model (dashed orange). The error in the observed ratio
    (gray ribbon) was computed by taking the standard deviation of the
    \hthcn{} and \hcfin{} fluxes at each ratio. The black horizontal bar
    shows the beam size of $0''.5$.}
  \label{fig:profile}
\end{figure}
}

\newcommand{\FigChannelMapsA}{%
\begin{figure}[b!]
  \centering
  \subfloat[\hthcn{}]{\includegraphics[width=0.95\textwidth]{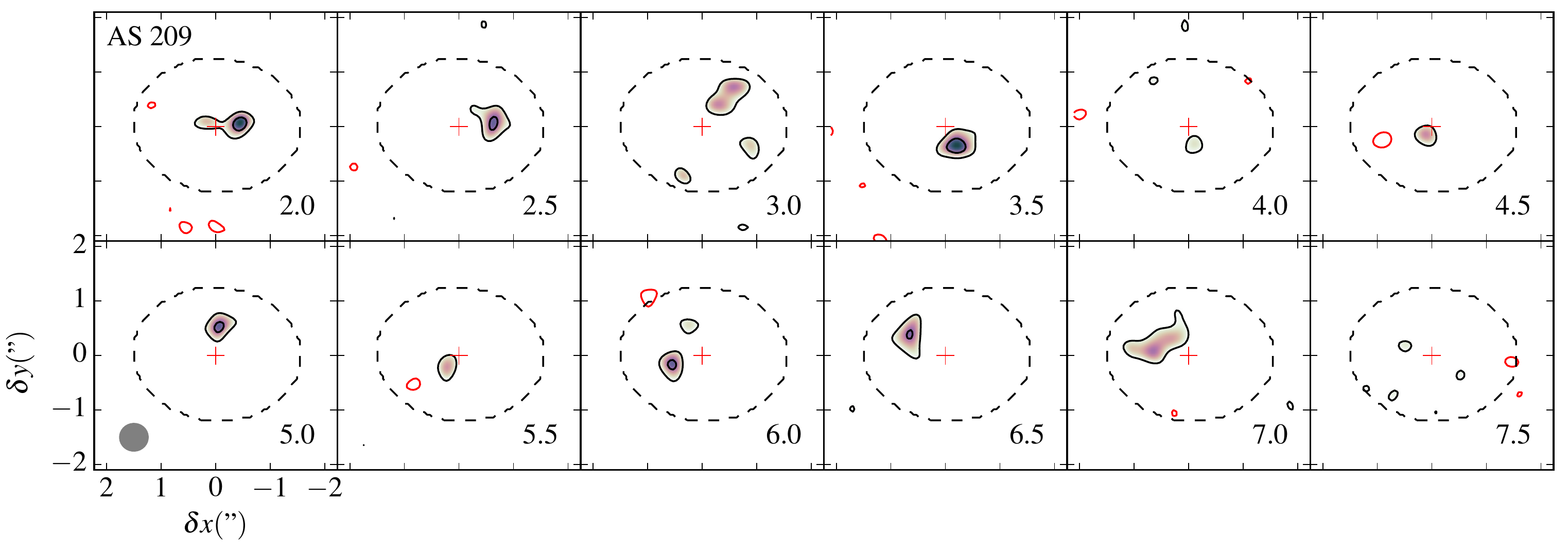}}\\
  \subfloat[\hcfin{}]{\includegraphics[width=0.95\textwidth]{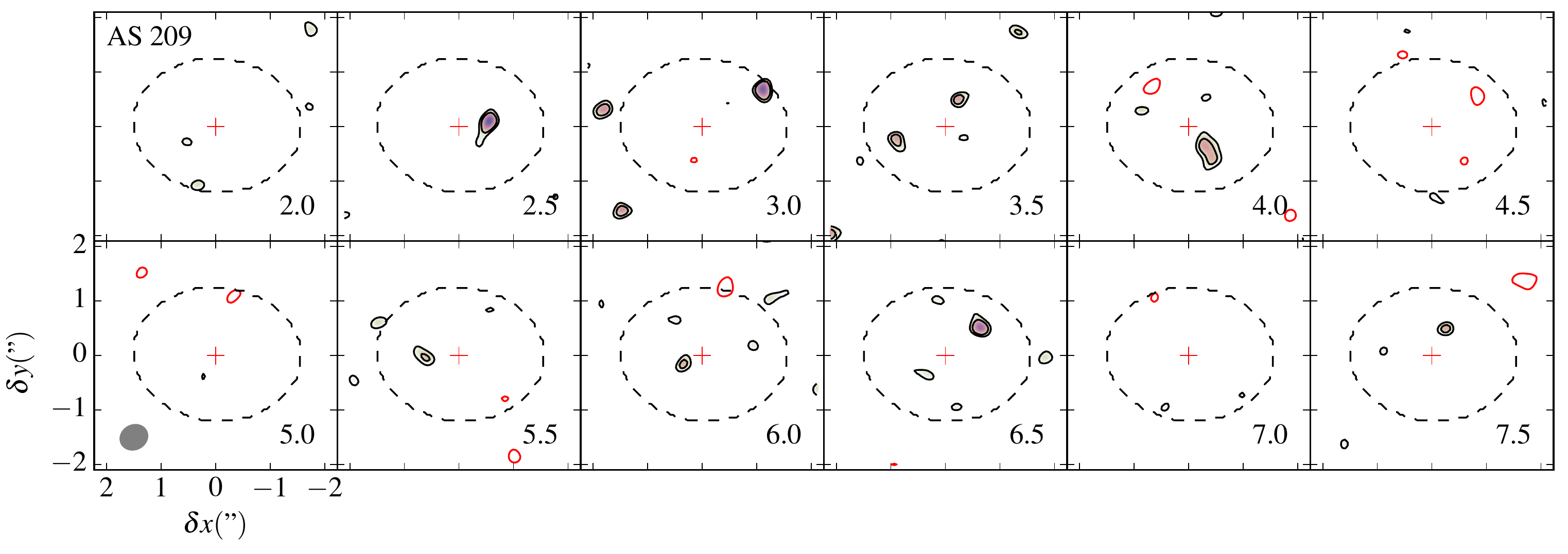}}
  \caption{Channel maps of the observed \hthcn{} (upper) and \hcfin{}
    (lower) $J=3-2$ emission lines in the disk around star AS~209. The
    contour levels correspond to 3, 5, 7 10, 15, 20 and 25$\times$rms,
    with rms given in Table~\ref{tab:obs}. Positive and negative
    contours are drawn in black and red, respectively. For \hcfin{},
  an additional contour corresponding to 2.5$\times$rms is drawn. The
  elliptical mask used to extract the spectra and integrated flux is
  overlaid in dashed lines. The red crosses mark the continuum image
  centroid. The synthesized beam is shown in the bottom left panels.}
  \label{fig:channel-maps-a}
\end{figure}
}

\newcommand{\FigChannelMapsB}{%
\begin{figure}[t!]
  \centering
  \subfloat[\hthcn{}]{\includegraphics[width=0.95\textwidth]{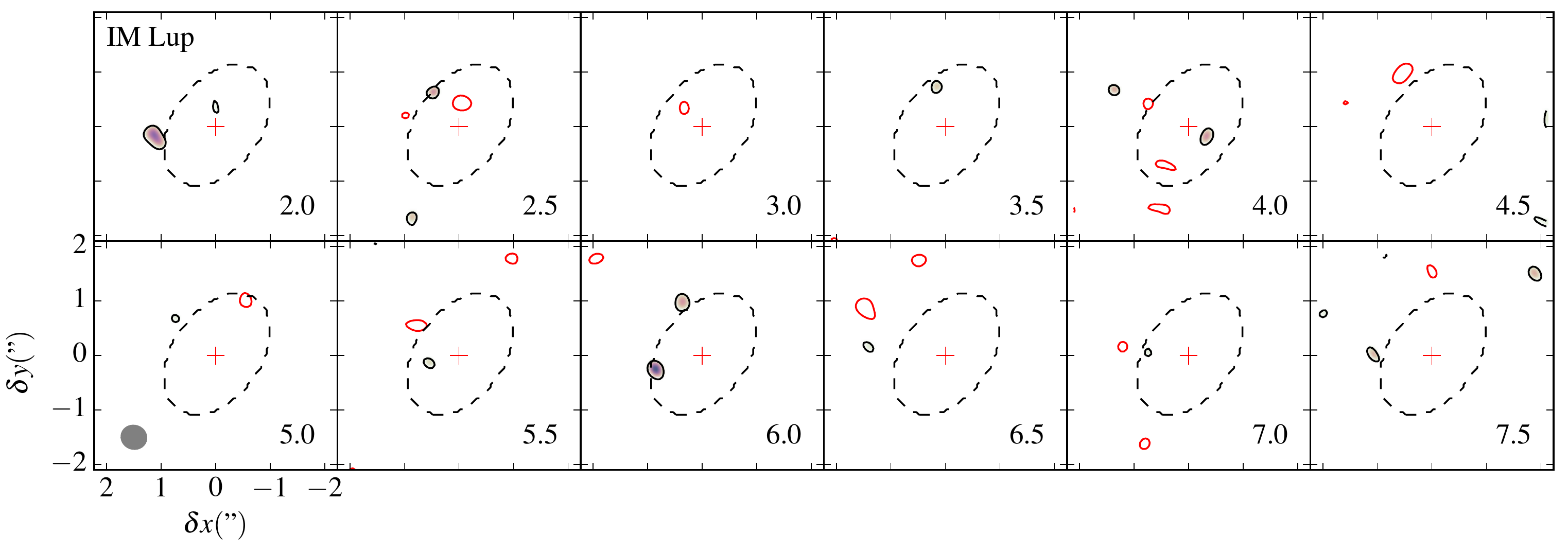}}\\
  \subfloat[\hcfin{}]{\includegraphics[width=0.95\textwidth]{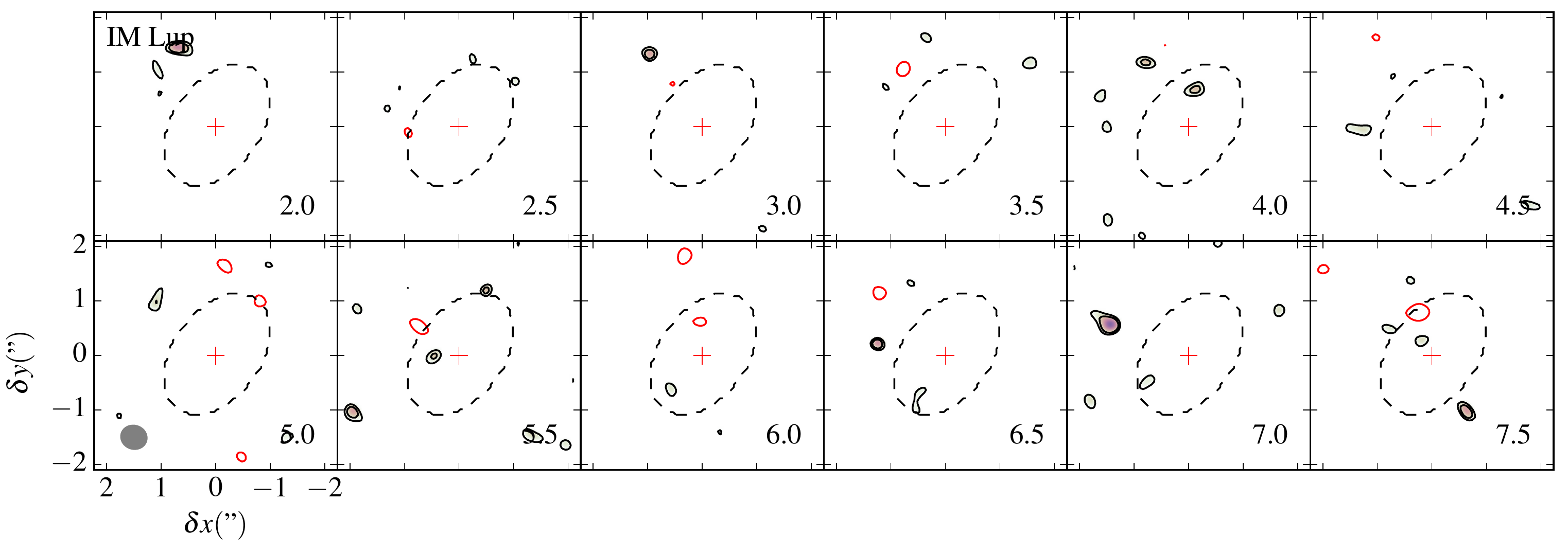}}
  \caption{Same as in Fig.~\ref{fig:channel-maps-a} but for IM~Lup.}
  \label{fig:channel-maps-b}
\end{figure}
}
\newcommand{\FigChannelMapsC}{%
\begin{figure}[t!]
  \centering
  \subfloat[\hthcn{}]{\includegraphics[width=0.95\textwidth]{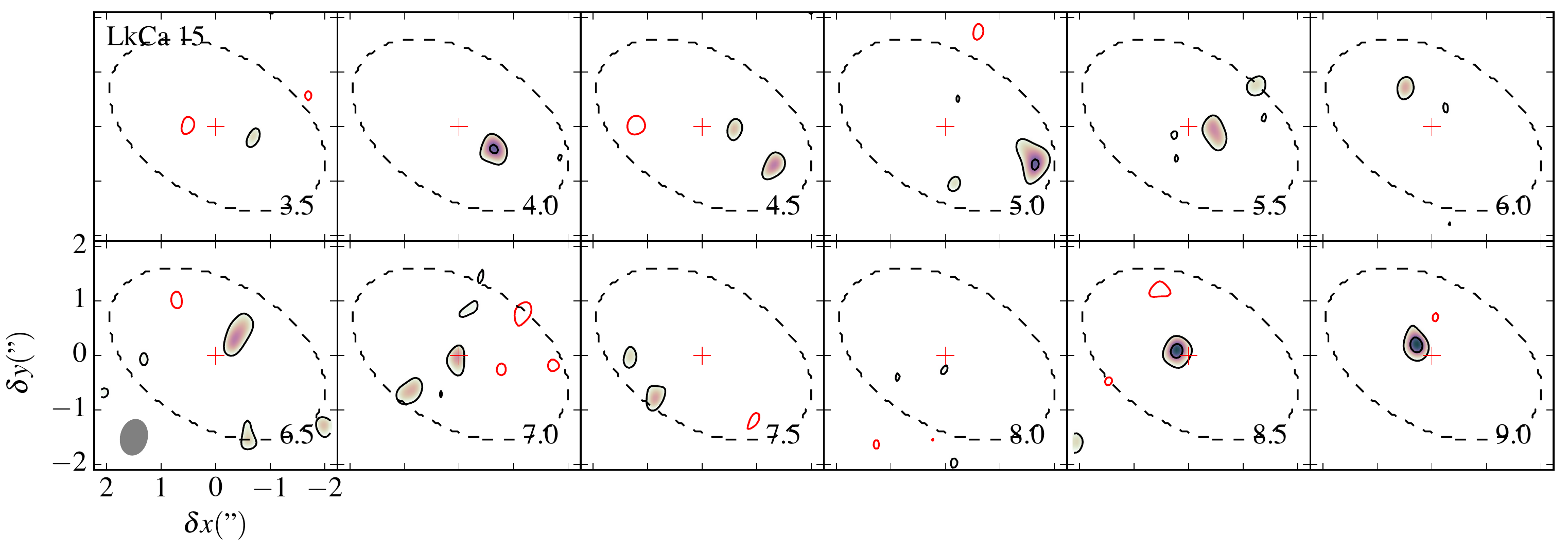}}\\
  \subfloat[\hcfin{}]{\includegraphics[width=0.95\textwidth]{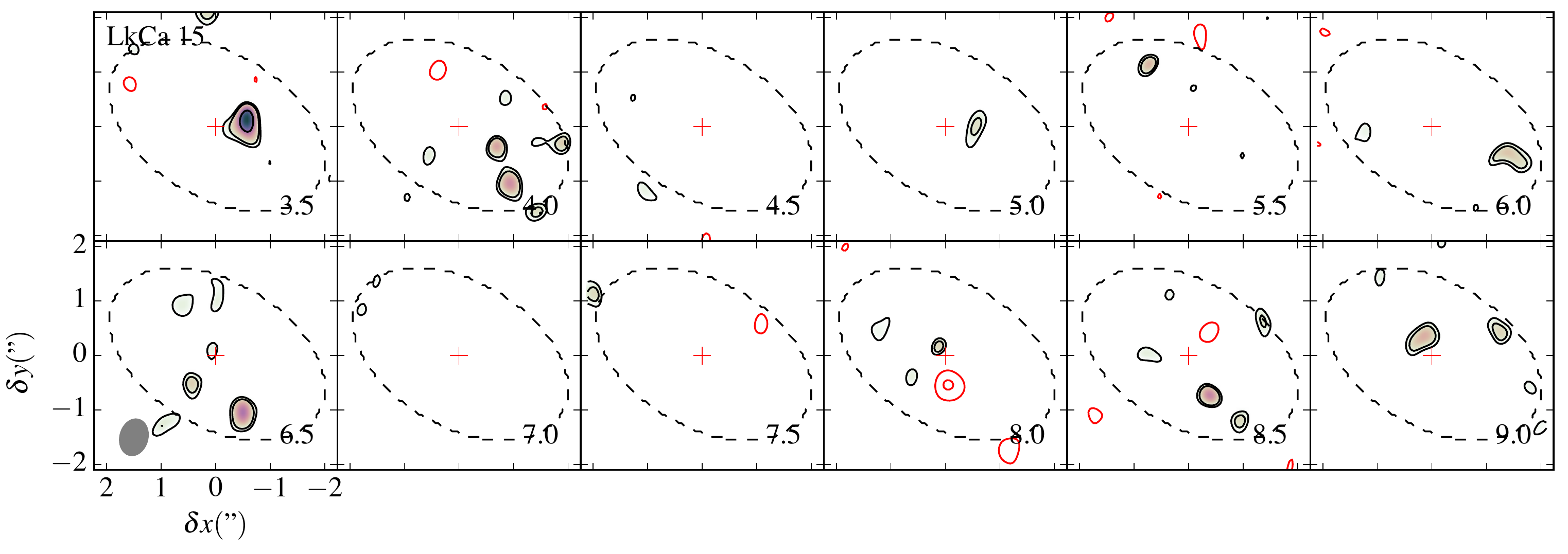}}
  \caption{Same as in Fig.~\ref{fig:channel-maps-a} but for LkCa~15.}
  \label{fig:channel-maps-c}
\end{figure}
}
\newcommand{\FigChannelMapsD}{%
\begin{figure}[t!]
  \centering
  \subfloat[\hthcn{}]{\includegraphics[width=0.95\textwidth]{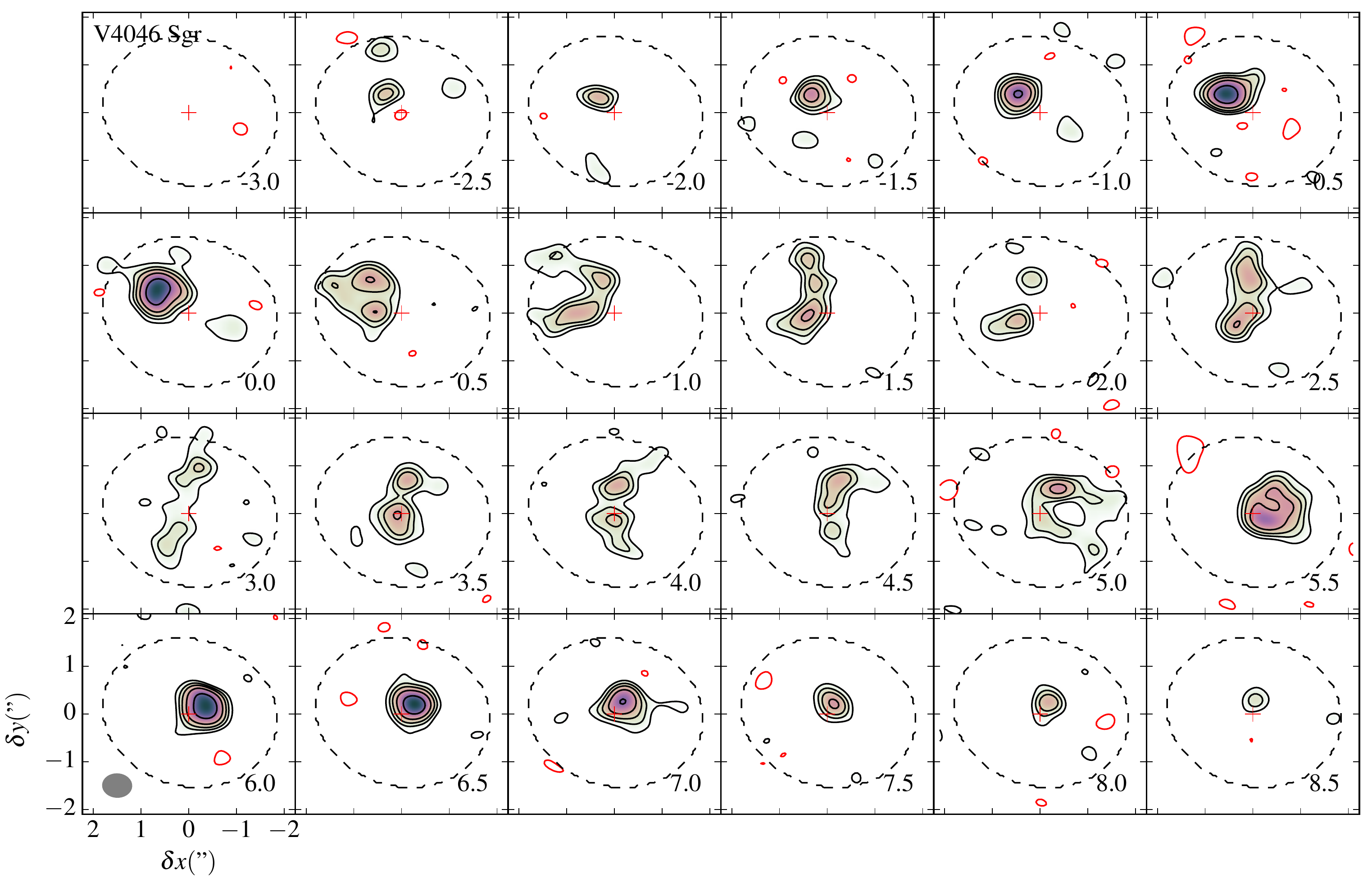}}\\
  \subfloat[\hcfin{}]{\includegraphics[width=0.95\textwidth]{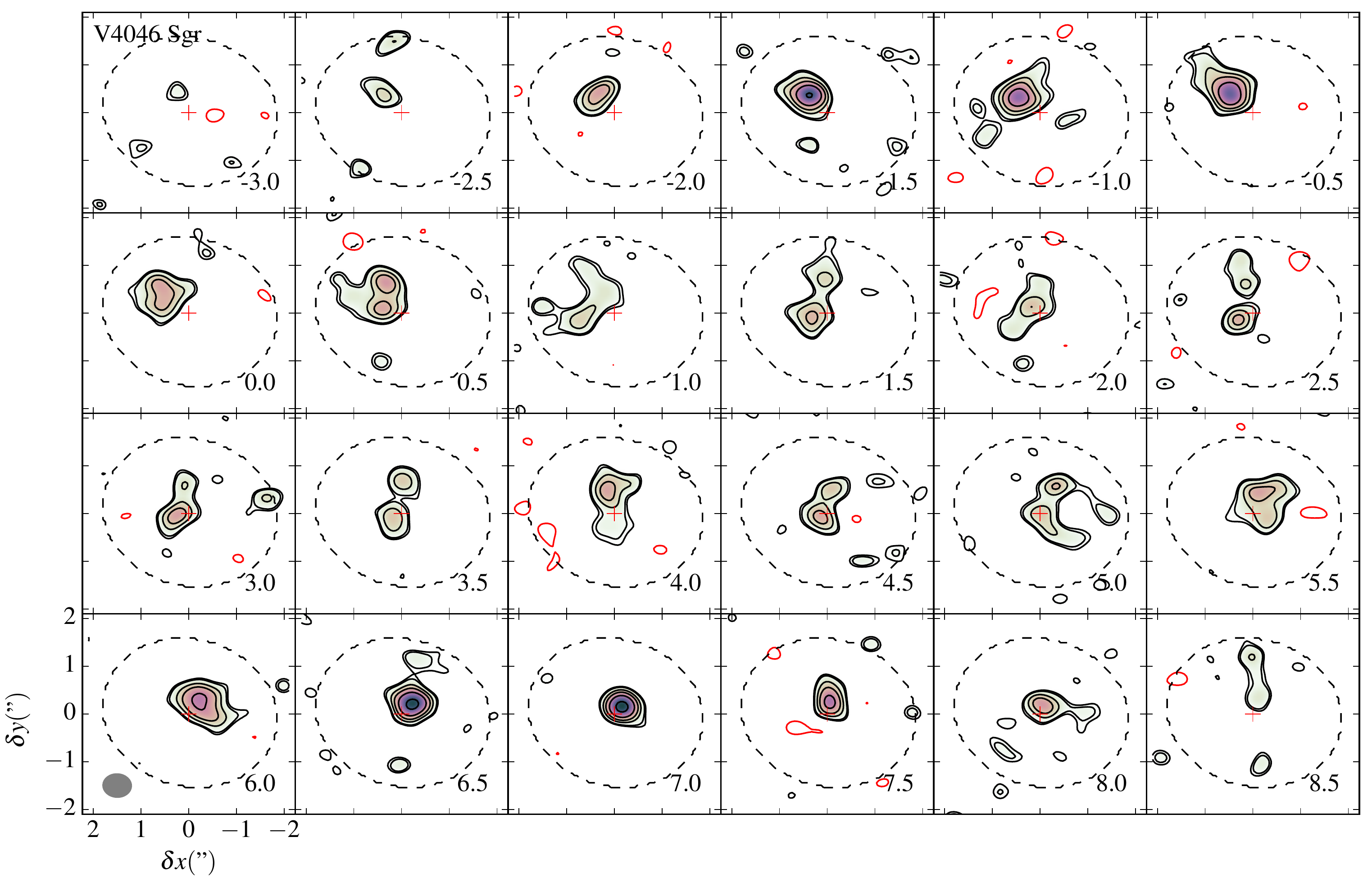}}
  \caption{Same as in Fig.~\ref{fig:channel-maps-a} but for V4046~Sgr.}
  \label{fig:channel-maps-d}
\end{figure}
}
\newcommand{\FigChannelMapsE}{%
\begin{figure}[t!]
  \centering
  \subfloat[\hthcn{}]{\includegraphics[width=0.95\textwidth]{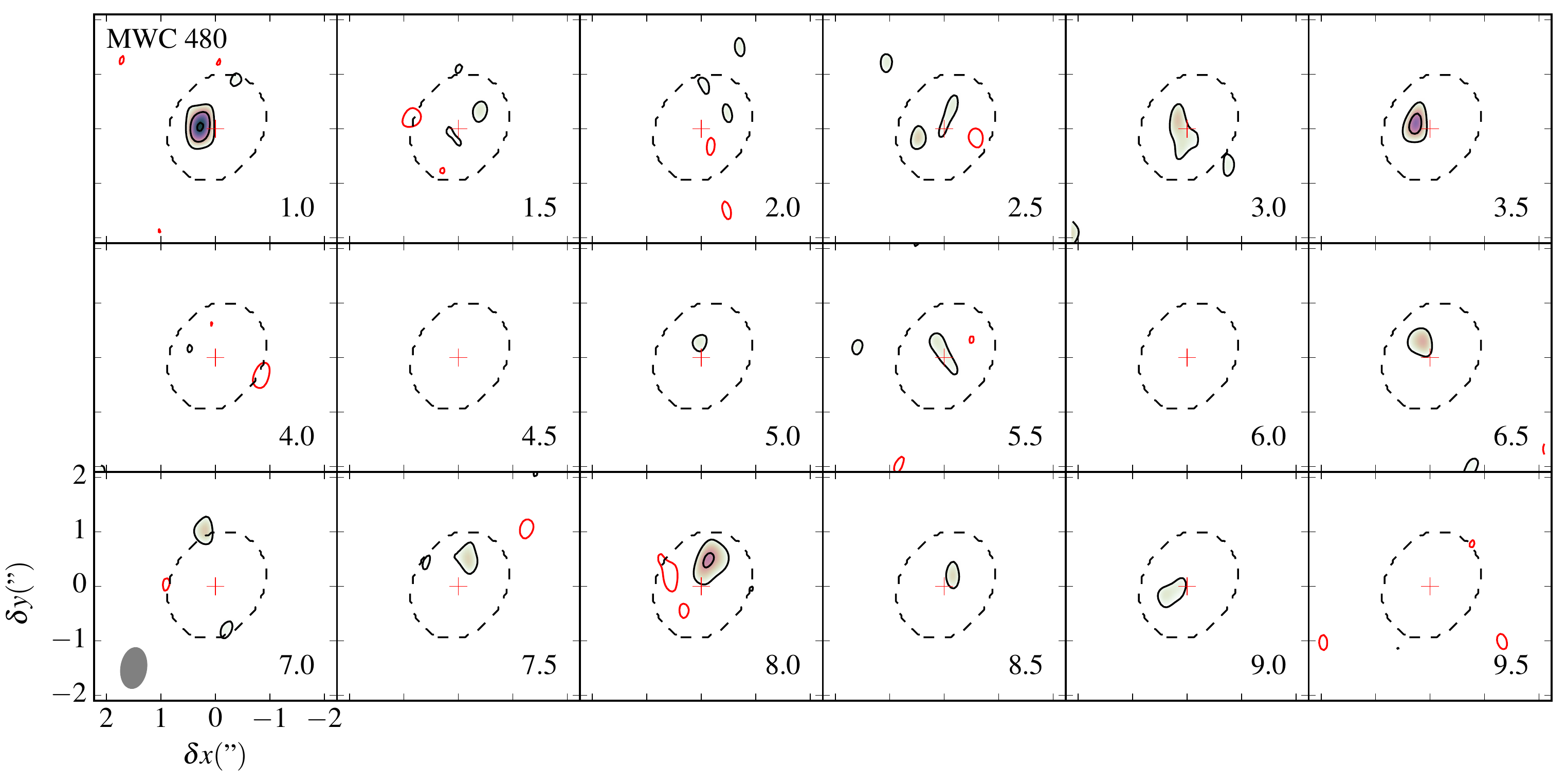}}\\
  \subfloat[\hcfin{}]{\includegraphics[width=0.95\textwidth]{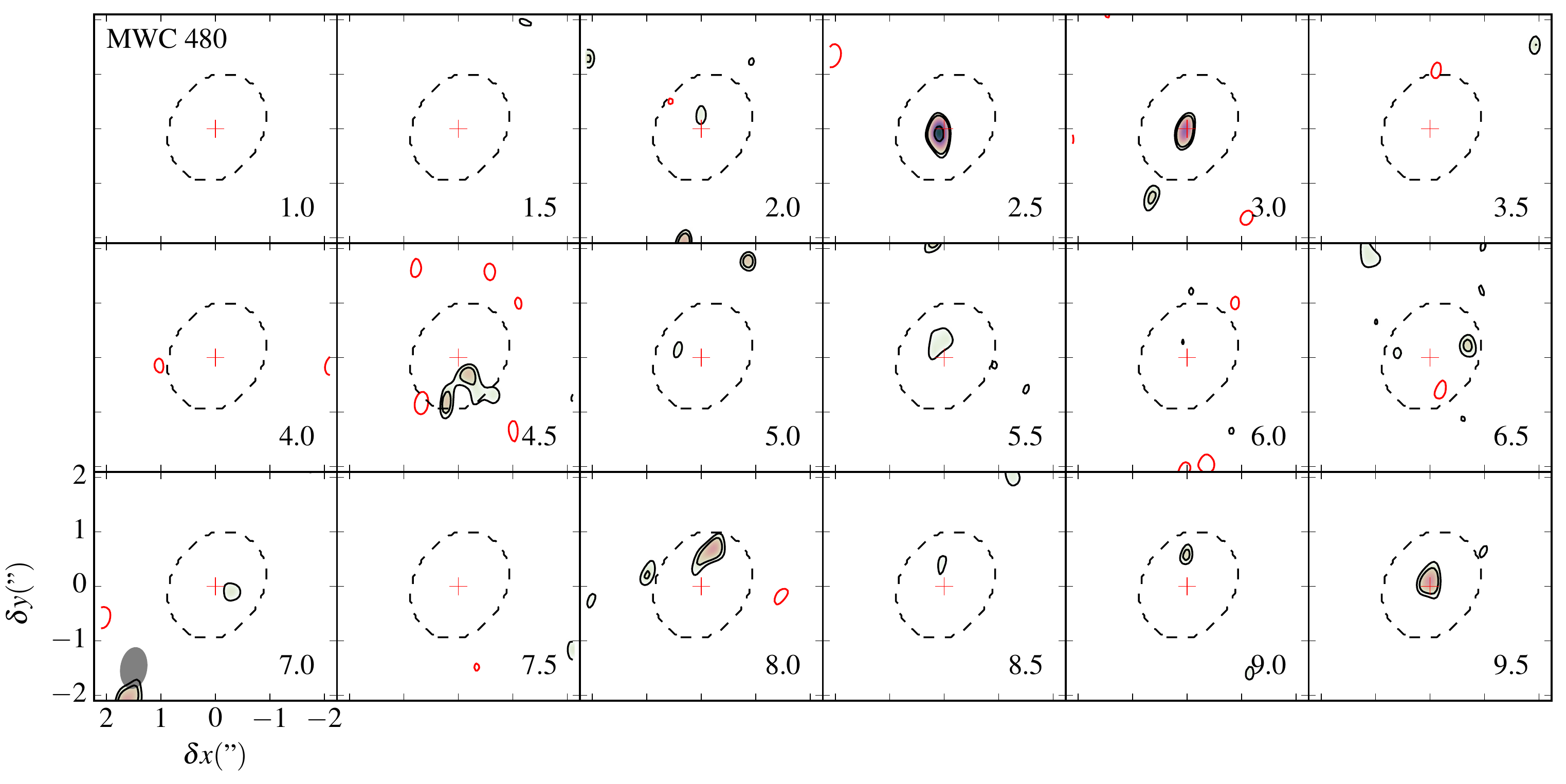}}
  \caption{Same as in Fig.~\ref{fig:channel-maps-a} but for MWC~480.}
 \label{fig:channel-maps-f}
\end{figure}
}
\newcommand{\FigChannelMapsF}{%
\begin{figure}[t!]
  \centering
  \subfloat[\hthcn{}]{\includegraphics[width=0.95\textwidth]{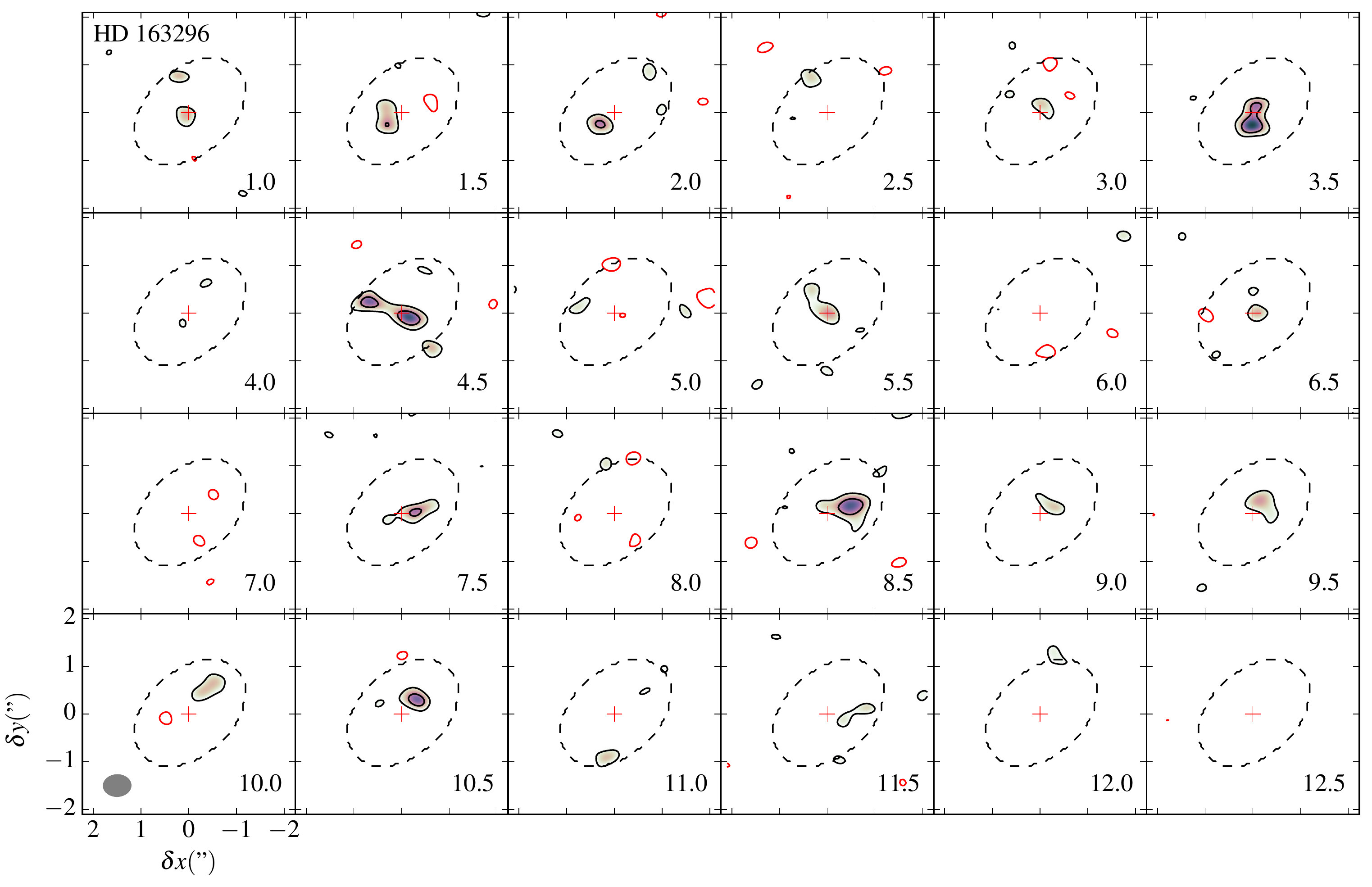}}\\
  \subfloat[\hcfin{}]{\includegraphics[width=0.95\textwidth]{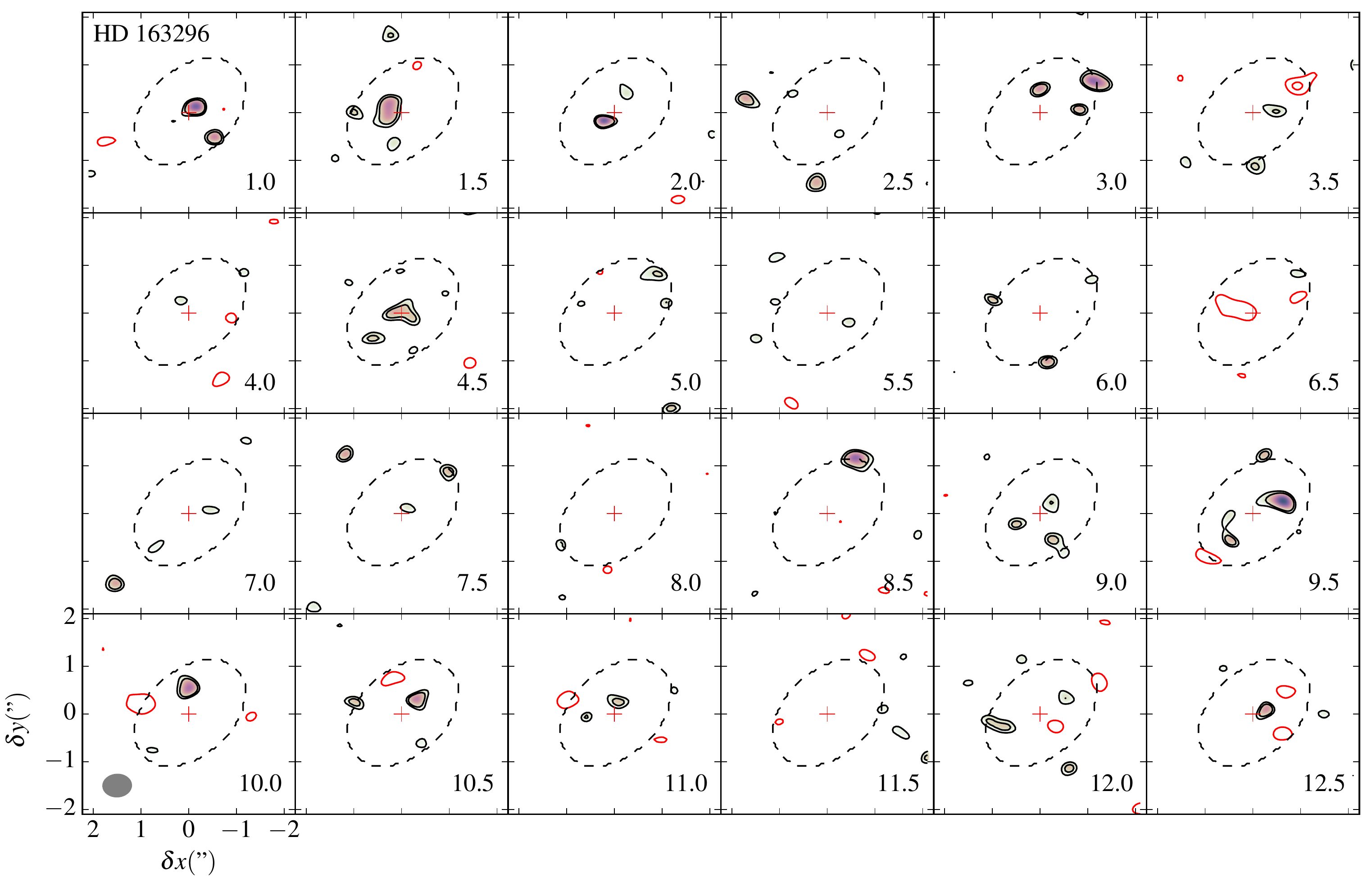}}
  \caption{Same as in Fig.~\ref{fig:channel-maps-a} but for HD~163296.}
  \label{fig:channel-maps-e}
\end{figure}
}


\section{Introduction}

\TabObs{}
\TabSources{}

The origin of Solar System organics is a fundamental and highly
debated topic. It is unclear whether the organics in the different
Solar System bodies were inherited from the cold and dense molecular
parent cloud of our Sun, or if they are the result of chemical
processing within the Solar nebula protoplanetary disk. The isotopic
composition of present day organics may help to shed light on their
origins, since isotopic fractionation chemistry is highly environment
specific and can leave a permanent imprint. Comets are especially
interesting in this context, since they should preserve the isotopic
compositions in different molecules during the assembly of the Solar
System.

Among the different methods used to trace the origin of molecules, the
\Nratio{} isotopic ratio is one of the most popular ones. Nitrogen
isotopic ratios span at least an order of magnitude between different
Solar System bodies. A low nitrogen fractionation (high \Nratio{}) is
observed toward the Sun and Jupiter \citep[$\Nratio=450$;
][]{marty2011}, while a high fractionation is observed in the rocky
planets, comets and meteorites
\citep[$\Nratio=100-300$;][]{mumma2011}. The origin of these \Nratio{}
variations is not well understood, but it suggests that different Solar
System bodies obtained their nitrogen from different nitrogen
reservoirs \citep{mumma2011}. Based on observations of the
Interstellar Medium (ISM), comets and chemical models, there are three
major nitrogen reservoirs in dense interstellar and circumstellar media,
N$_2$, NH$_3$ and HCN. These species have different fractionation
pathways \citep[\eg{},][]{hily-blant2013}, and it is thus important to
measure the \Nratio{} ratio in molecules representative of these nitrogen
reservoirs.

This study focuses on HCN. HCN is readily detected in the ISM
\citep[\eg{},][]{liszt2001,hily-blant2013}, comets
\citep[\eg{},][]{bockelee2008} and protoplanetary disks
\citep[\eg{},][]{thi2004,oberg2010,oberg2011,chapillon2012}. While several studies have
been made toward prestellar cores and protostars, the
characterization of isotopic ratios in protoplanetary disks is rather
new due to the intrinsic weak line emission. \citet{guzman2015}
presented the first detection of \hthcn{} and \hcfin{} in the disk
around Herbig Ae star MWC~480, and provided the first measurement of
the \Nratio{} in a disk. They found an isotopic ratio of $200\pm100$,
which is similar to what is observed in the cold ISM and in comets.

The low \Nratio{} value in the MWC~480 disk implies either inheritance
of HCN from the ISM, or the presence of an active fractionation
chemistry in the disk. There are two potentially active fractionation
channels in disks and in the ISM. The first is through isotope
exchange reactions, such as
\begin{equation}
\emr{HC^{14}NH}^+ + \emr{^{15}N} \rightarrow \emr{HC^{15}NH^+} + \emr{^{14}N} + h\nu
\end{equation}
which favor the incorporation of $^{15}$N into molecules at low
temperatures ($<20$~K). \emr{HC^{15}NH^+} can later recombine with
free electrons to produce \hcfin{}. Observations of HCN and HNC
fractionation toward protostars present a tentative trend of the
\Nratio{} with temperature, supporting this scenario
\citep{wampfler2014}. The second mechanism is selective
photo-dissociation of $^{14}$N$^{15}$N over $^{14}$N$_2$, due to
self-shielding of $^{14}$N$_2$ \citep{heays2014}. In the surface layers
of protoplanetary disks, which are directly illuminated by the
radiation field of the central star, the dominant formation pathways
leading to HCN and \hcfin{} are
\begin{equation}
\emr{^{14}N} + \emr{CH_2} \rightarrow \emr{HC^{14}N}
\end{equation}
\begin{equation}
\emr{^{15}N} + \emr{CH_2} \rightarrow \hcfin
\end{equation}
Both mechanisms can reduce the HCN/\hcfin{} ratio in protoplanetary
disks \citep{lyons2009,heays2014}. Distinguishing between these
different origins of nitrogen fractionation levels in disks and
between inheritance and in situ disk fractionation chemistry (and
further in comets and planets) requires more disk measurements, and
constraints on the radial profiles of the \Nratio{} ratio in disks
with different structures and around stars with different radiation
fields. A constant \Nratio{} ratio across disks would favor a scenario
where disks inherit their organics from the natal cloud, while disk
chemistry should result in a radial gradient, since the disk
environment is dramatically different at different radii.

To begin to address this long-term goal, we present observations of
\hthcn{} and \hcfin{} in a diverse sample of 6 protoplanetary
disks. Because the HCN lines may be optically thick, we use the
\hthcn{} line as a proxy of HCN to derive the \Nratio{} ratio. In
section~2 we present the observations and describe the data reduction
process. The disk-averaged isotopic flux ratios derived from the
observations are presented in Section~3. In Section~4, we model the
disk abundance profiles of \hthcn{} and \hcfin{} in V4046 Sgr, the
source with the highest signal-to-noise ratio detection. In Section~5,
we discuss the results and compare with observations in our Solar
System and in the cold ISM. A summary is presented in Section~6.

\FigSurvey{}

\section{Observations and data reduction}
\label{sec:obs}

The \hthcn{} and \hcfin{} $J=3-2$ lines were observed with ALMA during
Cycle 2 as part of project ADS/JAO.ALMA\#2013.1.00226. The Band~6
observations included two spectral settings, at 1.1 and 1.4~mm. The
correlator setup of the 1.1~mm and 1.4~mm settings were configured
with 14 and 13 narrow spectral windows, respectively, targeting
different molecular lines. The main targets of the observations were
deuterated species, which are presented in \citet{huang2017}. They include an independent analysis of the \hthcn{} data in
the context of D/H fractionation. A focused study on the DCO$^+$
emission in IM~Lup was presented earlier by \citet{oberg2015a}. The
\hthcn{} data in MWC~480 was also used for the study of CH$_3$CN by
\citet{oberg2015b}. In addition, \citet{guzman2015} presented the
\hthcn{} and \hcfin{} data in MWC~480. In this paper, we focus on the
\hthcn{} and \hcfin{} lines for the full sample. For consistency, we
present data reduction and imaged independently of the previous
papers.  For one of the sources, V4046~Sgr, we also use CO $J=2-1$
isotopologue data from the same survey.

The observations are described in detail by \citet{huang2017}. In
short, the Band 6 observations were carried-out between 2014 and 2015
with baseline lengths spanning between 18 and 650~m. The total
on-source time was 20~min, on average. A quasar was observed to
calibrate amplitude and phase temporal variations. To calibrate the
frequency bandpass a quasar was observed. The absolute flux scale was
derived by observing Titan for about half the observations, or a
quasar. The HCN isotopologue lines were covered by two spectral
windows of 59~MHz bandwidth and 61~kHz channel width in the 1.1~mm
spectral setting. The CO isotopologue lines were covered by three
spectral windows in the 1.4~mm spectral setting, with the same
bandwidth and channel width.
 
The data calibration was performed by the ALMA staff using standard
procedures. We took advantage of the bright continuum emission of the
sources to improve the signal-to-noise ratio, by further
self-calibrating the HCN isotopologue data. The self-calibration
solutions were derived on individual spectral windows when possible
(AS~209, LkCa~15, MWC~480 and V4046~Sgr) and on averaged spectral
windows for the weaker sources (IM~Lup and HD~163296), and then
applied to each spectral window. The continuum was then subtracted
from the visibilities to produce the spectral line cubes. The clean
images were obtained by deconvolving the visibilities in CASA, using
the CLEAN algorithm with Briggs weighting. The HCN isotopologue data
was regridded to a spectral resolution of $0.5\kms$ for the full
sample. The robust parameter was set to 1.0 for \hthcn{}, except for
IM~Lup where we used a value of 2.0. For \hcfin{} the robust parameter
was set to 2.0 to improve the signal-to-noise ratio, except for
V4046~Sgr and MWC~480 for which a value of 1.0 was used because of the
bright line emission. To help the cleaning process we created
elliptical masks, same for all channels, centered on the dust
continuum centroid, using the disk inclination and position angle
listed in Table~\ref{tab:sample}. The mask radius was chosen to cover
the stronger \hthcn{} line emission in all channels, or to cover the
dust continuum emission if the line was not \TabFluxes{} detected. The
resulting beam, channel rms and moment zero rms values are listed in
Table~\ref{tab:obs}. For the CO isotopologues the robust parameter was
set to 0.5, and a Keplerian mask was used to help the cleaning
process, created by selecting emission consistent with the expected
Keplerian rotation of the disk in each channel. 

\section{Sample statistics}   
\label{sec:results}

\TabSpecParam{}

The stellar and disk properties of the sample are summarized in
Table~\ref{tab:sample}. The sample includes 4 T~Tauri stars and 2
Herbig~Ae stars. The stellar masses range between 0.9 (AS~209) and
2.3~\Msun (HD~163296), corresponding to luminosities that span an
order of magnitude. Two of the sources, namely LkCa~15 and V4046~Sgr,
are transitional disks, with inner holes resolved at millimeter
wavelengths. The sample is biased toward large disks, with a known
rich molecular emission. However, given the very different physical
properties, in particular the gas temperature, the source selection
allows us to determine the disk averaged \Nratio{} ratio in HCN in a
diverse sample of disks.

Figure~\ref{fig:survey} shows the observations for the full sample of
disks. The dust continuum images are shown in the left column. The
1.1~mm continuum images were produced by averaging 1.1~mm line free
spectral windows. The four middle panels display the \hthcn{} and
\hcfin{} velocity integrated maps, for the full line (color images)
and for two velocity ranges, the blue and red shifted parts of the
line, to demonstrate the Keplerian rotation of the disk (blue and red
contours). The right column of the figure shows the disk integrated
line profiles of the \hthcn{} and \hcfin{} lines. The spectra were
extracted using the same elliptical masks used to clean the
data. Figs.~\ref{fig:channel-maps-a} to \ref{fig:channel-maps-e} show
channels maps of the \hthcn{} and \hcfin{} lines in each source, with
the elliptical masks overlaid on top.

The lines are classified as detected if emission consistent with the
expected Keplerian rotation of the disk is observed in at least three
channels at a $3\sigma$ level. From the inspection of the channel maps
we find that \hthcn{} is clearly detected toward all disks except for
the T Tauri disk IM Lup. \hcfin{} is clearly detected toward
V4046~Sgr, MWC~480 and HD~163296, weakly detected toward AS~209 and
LkCa~15, and not detected toward IM~Lup. We note that for the two
disks with weak \hcfin{} line emission, while emission is detected in 3
or more individual channels, the emission is almost washed out in the
integrated intensity maps.

The \hthcn{} emission is generally compact compared to the extent of
the dust disk. It appears centrally peaked toward MWC 480 and V4046
Sgr, and possibly toward HD 163296, but presents clear rings toward
the remaining two sources: LkCa 15 and AS 209. The latter is
unexpected, since AS 209 is not a transition disk.  The \hcfin{} line
shows a similar behavior, although the signal-to-noise ratio is lower.

We extracted the disk integrated fluxes from the unclipped moment zero
maps using the same elliptical mask created to clean the data and
extract the spectra. The uncertainty in the flux was estimated by
simulating integrated flux measurements from signal free regions using
the same elliptical mask but centered at random positions. The
integrated disk fluxes and their associated uncertainties are listed
in Table~\ref{tab:ratios}. 

We use the extracted line flux ratios to estimate the abundance ratios
of the two isotopologues. This is a reasonable first approximation
when comparing HCN isotopologes, as the \hthcn{} and \hcfin{} line
emission is expected to be optically thin, arise in the same region
and this region is dense enough for the molecular rotational
population to be in LTE \citep{pavlyuchenkov2007}. We compute the
$\hthcn/\hcfin$ abundance ratio in the LTE case and for
$\Tex=15$~K. For MWC~480, we obtain a lower $\hthcn/\hcfin$ ratio
($1.8\pm0.3$) than the value of $2.8\pm1.4$ reported in
\citet{guzman2015}, although both values are consistent within the
errors. This is due to the different method implemented in this paper
to extract the fluxes. The resulting $\hthcn/\hcfin$ abundance ratios
span from 1.2 to 2.2 with an average of 1.8. Given the almost
identical upper energies, Einstein coefficients and the ratio $Q/g_u$
(partition function over upper state degeneracy) for the $J=3-2$
transition (see Table~\ref{tab:spec}), the $\hthcn/\hcfin$ abundance
ratios are almost identical to the flux ratios. We note that a higher
excitation temperature of 100~K changes the inferred abundances by
less than 1\%.

\FigDNratio{}

In order to derive the nitrogen fractionation in HCN, we adopt an
isotopic ratio of $\Cratio=70$. Because the C isotopic ratio depends
on the physical conditions of the gas \citep[\eg{},][]{roueff2015} we
include a 30\% uncertainty in this value to convert the
\hthcn/\hcfin{} ratio into a \Nratio{} ratio. The inferred
\Nratio{} ratios span from 83 to 156 with an average of 124. All disks
present low cometary-like \Nratio{} ratios. The resulting
\hthcn/\hcfin{} abundance ratios and the inferred \Nratio{} flux
ratios are listed in Table~\ref{tab:ratios}. Fig.~\ref{fig:DNratios}
shows the nitrogen fractionation ratios for the full sample and
compares it with DCN/HCN ratios derived by \citet{huang2017}. There
is no indication of a correlation between the disk averaged nitrogen
and hydrogen fractionation in these disks, as might have been expected
if both originated from a cold fractionation pathway (see also
section~\ref{sec:disc:Nchemistry}). We note that the spread in
\Nratio{} is small compared to the errors and we cannot rule out that
there is a trend that is washed out by the noise.

\section{The \hthcn/\hcfin{} profile in V4046 Sgr}

HCN isotopologue emission observed toward V4046~Sgr is sufficiently
bright to provide constraints on the \hthcn{} and \hcfin{} abundance
profiles. In this section, we model the emission profile of the
observed lines in order to retrieve the underlying \Nratio{} abundance
ratio across the disk.

\subsection{Disk physical structure}

In order to investigate possible variations of the abundance ratio
across the disk a detailed model of the line emission is needed. We
build a parametric model to describe the physical structure of the
disk based on the model described in \citet{rosenfeld2013}, which was
constructed to reproduce the emission of the dust continuum and the CO
isotopologues. We first parametrize the dust surface density as
\begin{equation}
  \Sigma_{dust}(r) = 
\begin{cases}
  \Sigma_c \left (\frac{r}{r_c}\right)^{-\gamma} \exp
  \left [ -\left ( \frac{r}{r_c}\right )^{2-\gamma}\right ] & r \geq r_{cav}\\
  \Sigma_{cav} &  0.2\mathrm{AU}<r<r_{cav}\\
\end{cases}
\end{equation}
where $\Sigma_c$ is a normalization factor, $r_c$ is a characteristic
radius, $\gamma=1$ is the power-law index of the viscosity, and
$r_{cav}$ is the radius of the inner cavity. We include two dust
populations, one for the atmosphere and another for the midplane grains.
The dust volume density is computed assuming a vertical Gaussian distribution
of each dust grain population:
\begin{equation}
  \rho_{dust} = \sum_{i=0,1} \frac{\Sigma_i(r)}{\sqrt{2\pi} H_i(r)} \exp \left
  (\frac{-z^2}{H_i(r)^2} \right),
\end{equation}
The midplane grains, which are larger than the atmospheric ones,
comprise the bulk of the dust mass ($\Sigma_{mid} = 0.9
\Sigma_{dust}$, $\Sigma_{atm}=0.1 \Sigma_{dust}$). The atmospheric dust
grains are vertically Gaussian distributed with a scale height: 
\begin{equation}
H_{atm}(r) = H_{10} \left ( \frac{r}{100~{\rm AU}} \right )^h
\end{equation}
and the midplane dust grains are concentrated closer to the midplane,
with a scale height that is half of that of the atmospheric grains.
\begin{equation}
H_{mid}(r) = \frac{1}{2} H_{atm}(r)
\end{equation}
With the dust density described above, the radiative transfer code
RADMC-3D \citep{dullemond2012} was then used to compute the dust
temperature throughout the disk. The dust absorption and scattering
opacities, which are needed to solve the thermal balance, were
computed using the {\texttt
  OpacityTool}\footnote{https://dianaproject.wp.st-andrews.ac.uk/data-results-downloads/fortran-package/}
from the DIANA project \citep{woitke2016}. The code assumes a mixture
of amorphous laboratory silicates with amorphous carbon and 25\%
porosity for the grain composition. The grain size distribution
follows a power-law of index $-3.5$. The minimum grain size was set to
5~nm, and the maximum size was set to 10~$\mu$m and 1~cm, for the
atmosphere and midplane grain populations, respectively.

\TabParam{}

The gas temperature is parametrized as
\begin{equation}
  T_{gas}(r,z) =
  \begin{cases}
  T_a + (T_m-T_a) \left ( \cos \left (\frac{\pi z}{2 z_q} \right) \right )^{2\delta} & z<z_q\\
  T_a & z\geq z_q
  \end{cases}
\end{equation}
following \citet{dartois2003}. Here, the atmospheric temperature is
given by a power-law ($T_a=T_{a,0} (r/10)^{q_{atm}}$), and the
midplane temperature is fixed to the dust temperature
($T_m=\Tdust(z=0)$). The fiducial scale height at which the gas
temperature is allowed to vary vertically, $z_q$, is fixed to $z_q=2
H_{gas}$, where $H_{gas}=2 c_s/\Omega$ is the hydrostatic gas scale
height evaluated at the midplane, $z=0$. 

Once the gas temperature is obtained, the hydrostatic equation is
solved to derive the gas density across the disk. For this we assume a
vertically integrated gas-to-dust ratio at each radii of 100. The
adopted parameters for the model are listed in
Table~\ref{tab:param}. The resulting gas density and temperature
structures are shown in Figure~\ref{fig:structure}. This model
reproduces the main features of the $^{12}$CO, $^{13}$CO and C$^{18}$O
emission (see Fig.~\ref{fig:COmodel}) well enough for the purpose of
this study, assuming standard isotopic ratios. In this model, the CO
abundance is kept constant throughout the disk, except in the cold
midplane ($\Tgas<19$~K) where the abundance is reduced by a factor of
$10^3$ due to freeze-out onto dust grains, and in the disk atmosphere
where the CO abundance is reduced by a factor $10^8$ due to
photodissociation.

\FigDiskStructure{}
\FigCOmodels{}

\subsection{Abundance fitting}

The molecular abundances for \hthcn{} and \hcfin{} were defined as
power-laws,
\begin{equation}
X=X_0(r/R_{0})^{\alpha}
\end{equation}
where $X_0$ is the abundance with respect to total hydrogen at
$R_0=100$~AU and $\alpha$ is a power-law index. We also include an
outer cut-off radius $\Rout$. This parametrization is a common
approach when modeling molecular abundances in disks
\citep[\eg{}][]{qi2008,qi2013}.

In order to find the model that best reproduces the HCN isotopologue
observations and the associated uncertainties, we use a Bayesian
approach. In short, we first create a synthetic observation of the
line emission for each species separately. Taking advantage of the
bright HCN isotopologue emission in V4046~Sgr, we produced observed
visibilities and cleaned cubes at a higher spectral resolution of
$0.2\kms$ for the line modeling and include include 60 channels. We
use the {\texttt vis\_sample} Python
package\footnote{https://pypi.python.org/pypi/vis\_sample} to compute
the Fourier Transform of the synthetic model and obtain visibilities
that are correctly re-projected on the $u-v$ points of the
observations. The likelihood function is then computed in the $u-v$
plane, by computing the weighted difference between model and
observations, for the real and imaginary parts of the complex
visibility. We sample the posterior distribution with the MCMC method
implemented in the {\texttt emcee} package by \citet{{foreman2013}}.

We include two free parameters in the line modeling, that is $X_0$,
$\alpha$, which are associated with the molecular abundances of \hthcn{}
and \hcfin{}. The outer radius, $R_{out}$, was fixed to 100~AU (chosen
by the extension of the emission in the moment-zero map), but we
checked that a larger radius of 200~AU gave the same result. The disk
physical structure, that is the gas density and temperature, are fixed
in the line fitting. We adopt the disk inclination, position angle,
stellar mass (including both stars) and systemic velocity listed in
Table~\ref{tab:sample}. The level populations were computed using
RADMC and assuming the gas is under LTE. We checked that non-LTE
effects are not important for these lines using the non-LTE radiative
transfer code LIME \citep{brinch2010} to re-calculate the level
populations for the best-fit model. When generating a new sample, we
included a flat prior for the power-law index $-3<\alpha<2$, and for
the molecular abundance $10^{-20}<X_{100}<10^{-8}$.

\FigRadialProfile{}

\FigTriangles{}
\FigResiduals{}

The best-fit model corresponds to $X_0=8.94\pm0.30\times10^{-13}$ and
$\alpha=-0.69\pm0.03$ for \hthcn{}, and
$X_0=3.37\pm0.19\times10^{-13}$ and $\alpha=-1.08\pm0.04$ for
\hcfin{}. Our model suggests an increasing $\hthcn/\hcfin$ ratio as a
function of radius, \ie{} higher fractionation in the inner disk
compared to the outer disk. Fig.~\ref{fig:profile} shows the
deprojected radial profiles of the dust continuum and HCN isotopologue
emission in V4046~Sgr (left panel) as well as the observed and modeled
$\hthcn/\hcfin$ flux ratio (right panel). We note that beyond 60~AU,
the ratio is highly uncertain because the signal-to-noise ratio
becomes too low, in particular for \hcfin{}. Figure~\ref{fig:mcmc}
shows the posterior probability distribution for the fitted
parameters. Fig.~\ref{fig:residuals} shows the residuals between our
best-fit model and the data for selected channels. Our simple model
is able to well reproduce the observations.

The inferred abundance $\hthcn/\hcfin$ ratio at 10 and 50~AU are
$1.08\pm0.14$ and $2.02\pm0.27$, respectively. Assuming
$\Cratio=70\pm21$, we infer an abundance \Nratio{} ratio of $76\pm25$
and $142\pm47$, at 10 and 50~AU respectively.  
We note that the derived disk integrated HCN/\hcfin{} flux ratio of
$115\pm43$ falls in between the inferred abundance ratios at 10 and
50~AU. Given the consistent nitrogen fractionation levels inferred from the
observations and line modeling in V4046~Sgr, we expect the observed
\Nratio{} ratio in the other sources to be representative of their
abundance ratio in the comet forming region.


\section{Discussion}
\label{sec:discussion}

\subsection{Disk-averaged nitrogen fractionation in protoplanetary disks}


We have shown that HCN isotopologues are abundant in disks. Both
\hthcn{} and \hcfin{} are detected toward 5/6 disks in our sample --
the one exception being the disk around T~Tauri star IM~Lup. The disk
around IM~Lup is very massive ($\Mdisk=0.1\Msun$), very cold and also
very young \citep[\eg{},][]{cleeves2016} compared to the rest of the
disks in the sample. The non-detection of \hthcn{} was surprising
considering that IM Lup is quite bright in the main HCN isotopologue
\citep{oberg2011}. Given the observed HCN flux density of
3.5~Jy~\kms{}, we could expect a \hthcn{} flux density of
50~mJy~\kms{} if HCN is optically thin and $\Cratio=70$. This is
consistent with the observed $3\sigma$ upper limit of $51$~mJy~\kms{}.

The inferred disk-averaged nitrogen fractionation ratios range from $83\pm37$
to $156\pm78$. Despite the different physical conditions of the disks
in the sample, the observed $\hthcn/\hcfin$ ratios are consistent with
sampling a constant disk-averaged fractionation level. In particular,
we find no difference in the nitrogen fractionation level between disks
around T~Tauri and Herbig~Ae stars, which have an order of magnitude
difference in the stellar radiation field. No difference in the
disk-averaged \Nratio{} is observed between full and transitional
disks, either. The age of the star does not seem to play an important
role either -- we target young ($\sim$1~Myr) and old ($>10$~Myr)
sources -- suggesting either that the \Nratio{} is inherited from the
parent cloud and is not modified in the disk, or the disk chemistry
sets the global, disk averaged nitrogen fractionation level early
($\lesssim1$~Myr) in the protoplanetary disk life. 

Although we do not observe differences in the nitrogen fractionation ratio
between the sources, the data suggest that there may be a difference
in the nitrogen abundance between the disks around stars V4046~Sgr,
MWC~480 and HD~163206, and the disks around T~Tauri stars AS~209 and
LkCa~15. The old disk around the binary T~Tauri stars V4046~Sgr and
the two Herbig Ae stars in the sample are enriched in HCN compared to
the young T~Tauri stars. Disk models have shown that dust migration
and carbon and oxygen depletion (mainly due to CO and H$_2$O
freeze-out) can increase the column density of cyanides by up to two
orders of magnitude in the outer disk \citep{du2015}. Future
observations toward a larger sample of disks will show if this
corresponds to an evolutionary trend.

\subsection{Comparison between nitrogen fractionation in disks, Solar System and ISM}

Fig.~\ref{fig:Nratios} shows the observed \Nratio{} ratios in
different Solar System bodies, the cold ISM and in the diffuse medium.
There are large variations in the \Nratio{} ratio among different
Solar system bodies, in particular between the rocky and gaseous
bodies. The Sun has the highest \Nratio{} value in this comparison
($441\pm5$), measured by the Genesis mission that sampled solar wind
ions, N$^+$ among them \citep{marty2011}. An almost identical value
was found in the atmosphere of Jupiter through the NH$_3$ observations
carried-out by the Cassini spacecraft \citep{fouchet2004}. Both
measurements are expected to trace the lack of nitrogen fractionation
in N$_2$, the main nitrogen reservoir of the protosolar nebula. The
$^{15}$N-depleted Solar value is thus considered to be representative
of the conditions of the gas when the Sun formed. All the other Solar
system bodies are enriched in $^{15}$N compared to the Sun. A value of
$\Nratio=272$ is found in the Earth's atmosphere, measured in
N$_2$. The \Nratio{} has also been measured in several comets. An
isotopic ratio of $\sim150$ was found in C/1995 O1 (Hale-Bopp) and
17P/Holmes, value which was consistent for both HCN and CN
\citep{bockelee2008}. Observations of 18 comets from both the Oort
cloud and the Kuiper Belt all show consistently low
HCN$/\hcfin\simeq100-250$ ratios \citep{mumma2011}.

\FigIsotopologuesRatios{}

The cold interstellar medium is also enriched in $^{15}$N.
\citet{hily-blant2013} measured the $\hthcn/\hcfin$ toward two
prestellar cores, L183 and L1544, and found \Nratio{} values of
$140-250$ and $140-360$, respectively. The \Nratio{} in CN was later
measured toward L1544 resulting in a surprisingly high CN/C$^{15}$N
ratio of $500\pm75$ \citep{hily-blant2013b}. The authors were able to
reproduce the observed difference in CN and HCN with chemical models
of cold gas. The fact that CN and HCN present similar fractionation
levels in comets, could be explained if CN is produced in the coma
from photo-dissociation of HCN
\citep{hily-blant2013b}. \citet{ikeda2002} also found a low
HCN$/\hcfin$ ratio of $151\pm16$ toward the prestellar core
L1521E.

Finally, the averaged $^{15}$N enrichment observed for HCN in
protoplanetary disks are similar to comets and the cold ISM. While the
similar nitrogen fractionation ratios found in the cold ISM, comets
and disks is consistent with an inheritance scenario for the origin of
organics in the Solar System, as we discuss in the next section, the
increasing \Nratio{} ratio in the disk of V4046~Sgr suggests that in
situ disk chemistry also contributes to the observed fractionation
patterns.


\subsection{Resolved nitrogen fractionation chemistry in disks}
\label{sec:disc:Nchemistry}

The bright emission of the HCN isotopologues in V4046~Sgr allows us to
trace the \Nratio{} profile across a disk for the first time. In
general, there are three possibilities for what could be observed: a
flat, decreasing or increasing \Nratio{} as a function of disk
radius. If \hthcn{} and \hcfin{} are inherited from the prestellar gas
and no further chemical processing occurs in the disk, then a constant
\Nratio{} ratio is expected across the disk. On the other hand, if the
nitrogen chemistry is altered by in-situ chemical fractionation in the
disk then a varying \Nratio{} is expected. In this case, if chemical
fractionation dominates the nitrogen fractionation then a low
\Nratio{} is expected to occur in the outer disk, where the gas
temperature is low (\ie{}, a decreasing \Nratio{} with radius). In
contrast, if selective photodissociation is the dominant pathway to
fractionate HCN, then an increasing \Nratio{} profile would be
observed because this pathway is most important in regions exposed to
UV photons, \ie{} the inner disk which is illuminated by the central
star.

The observations toward V4046~Sgr show that both \hthcn{} and
\hcfin{} are best reproduced by a decreasing abundance
profile. However, the inferred \hcfin{} emission profile is slightly
steeper than that of \hthcn{}, pointing to a higher fractionation in
the inner disk than in the outer disk. The varying \Nratio{} observed
in V4046~Sgr shows that there is an active nitrogen fractionation in
the disk, that changes the original fractionation pattern. The fact
that \Nratio{} is lower in the inner disk suggests that selective
photodissociation is indeed an important pathway to fractionate HCN in
the inner disk. Higher signal-to-noise ratio observations are needed
to determine if there is also an active N fractionation chemistry in
the outer disk.

There is additional evidence that cold ion-molecule fractionation does
not alone regulate the $\hthcn/\hcfin$ ratio in
disks. \citet{huang2017} measured the D/H isotopic ratio in DCO$^+$
and HCN toward the same sample of disks. They found enhanced D/H
ratios compared to the elemental ratio in the local ISM
($\sim2\times10^{-5}$) in all disks, with DCN/HCN ratios ranging from
0.005-0.08. If the cold pathway dominates the fractionation for both
species in these disks we could expect a correlation between the D/H
and \Nratio{} ratios. However, we do not see such correlation in the
sample (see Fig.~\ref{fig:DNratios}).

\subsection{Future directions}

We have shown that the \Nratio{} ratio increases with radius in the
disk around T~Tauri stars V4046~Sgr. However, the angular resolution
of the current data ($0''.5$ corresponding to $\sim36$~AU at 73~pc)
prevents us from resolving \Nratio{} variations at smaller scales and
the low signal-to-noise ratio prevents us from constraining the
chemistry beyond 60~AU. Future observations at higher angular
resolution should allow us to measure the radial dependence of
\Nratio{} at Solar System scales, and determine how the fractionation
ratio changes from the inner ($<15$~AU) to the outer ($>30$~AU) disk.

High-angular resolution observations toward more disks are needed to
determine whether an increasing \Nratio{} is a general characteristic
of disks or unique to V4046~Sgr. A larger sample is also needed to
draw more general conclusions on the dominant fractionation pathways
in disks, and how the \Nratio{} pattern depends on the physical
conditions of the disk. In this respect, new chemical models that
include all nitrogen fractionation pathways (selective
photodissociation and cold isotope exchange reactions) as well as
inheritance from the parent cloud are key to interpret the
observations. In addition, it is desirable to include the
fractionation of both carbon and nitrogen in the models, since most
\Nratio{} measurements in both disks and cold ISM rely on observations
of the $^{13}$C isotopologues to measure the contribution of the main
isotopologues.

In contrast to HCN, hydrides (\eg{}, N$_2$H$^+$ and NH$_3$) are found
to be $^{15}$N-depleted in the ISM. Toward L1544,
\cite{bizzocchi2013} found a nitrogen isotopic ratio in N$_2$H$^+$ of
1000. Toward the class 0 protostar B1b, \citet{daniel2013} found
\Nratio{} ratios of $260-355$ for NH$_3$ and an upper limit of $>600$
for N$_2$H$^+$. \citet{hily-blant2013} proposed that the difference in
nitrogen fractionation between cyanides and hydrides in the cold ISM
is the result of their different chemical origins: HCN derives from
atomic N, while NH$_3$ and N$_2$H$^+$ derive from molecular
nitrogen. This hypothesis, however, is challenged by the recent
measurement of the \Nratio{} in NH$_2$, a photodissociation product of
NH$_3$ in cometary coma, toward comet C/2012~S1 (ISON), where a
fractionation ratio of $139\pm38$ was found \citep{shinnaka2014}. A
similarly low value ($\sim130$) was found previously by
\citet{rousselot2014} based on the averaged spectrum of 12
comets. Toward comet ISON, CN and HCN were also found to be highly
enriched in $^{15}$N ($\Nratio\sim150$). As explained by
\citet{shinnaka2014}, one possibility to obtain similar HCN and NH$_3$
fractionation levels in comets, despite very different fractionation
levels in the cold ISM, is through grain surface chemistry. Indeed,
the ISM values represent the \Nratio{} in the gas-phase, while
cometary values represent the \Nratio{} in the ices. It is also
possible that the measured cometary values is not representative of
the composition of the nucleus. Measurements of the isotopic ratio in
NH$_3$ and N$_2$H$^+$ in disks would provide additional clues to
answer these questions. In particular, if the cyanide/hydride
dichotomy observed in the cold ISM holds for disks as well.

Finally, more observations toward comets targeting different
molecular {\it parent} species (\ie{}, species tracing the cometary
nucleus and not daughter species which are produced in the coma) will
be important to compare with observations in the ISM and disks, and to
elucidate the origins of $^{15}$N enhancements, and ultimately the
origins of organics, across the Solar System.

\section{Conclusions}
\label{sec:conclusions}

We have presented ALMA observations at $\sim0''.5$ angular resolution
of the HCN isotopologues in a diverse sample of six protoplanetary
disks. The sample contains 4 T~Tauri and 2 Herbig~Ae stars, which
sample an order of magnitude in radiation fields. Both \hthcn{} and
\hcfin{} are detected toward all the sources, except for IM~Lup,
which is the most massive and coldest disk, and likely the youngest,
in the sample. Adopting a standard \Cratio{} ratio of 70 (with a 30\%
uncertainty), we infer disk-averaged \Nratio{} ratios of $80-160$ for
the sources. Despite the different physical conditions of the disks,
\Nratio{} in HCN is similar for all sources. No differences are
observed between T~Tauri and Herbig~Ae stars, or between the full
disks and the transitional disks, which feature large dust
cavities. Also, no correlation is observed between disk-averaged D/H
and \Nratio{} ratios in the sample. The observed disk-averaged
\Nratio{} ratios are similar to what is observed in comets and in the
cold ISM, which is consistent with the inheritance scenario for the
origin of organics in the Solar System. However, chemical processing
within the protoplanetary disk phase based on these ratios alone
cannot be ruled out. Indeed, in the one disk where we could resolve
the $\hthcn/\hcfin$ ratio as a function of radius we find a slightly
steeper profile for \hcfin{}, \ie{} an increasing \Nratio{} ratio with
radius. The higher nitrogen fractionation level in the inner disk
compared to the outer disk suggest that selective photodissociation is
an important fractionation pathway in the inner disk.

\begin{acknowledgments}
This paper makes use of ALMA data, project code:
ADS/JAO.ALMA\#2013.1.00226. ALMA is a partnership of ESO (representing
its member states), NSF (USA) and NINS (Japan), together with NRC
(Canada) and NSC and ASIAA (Taiwan), in cooperation with the Republic
of Chile. The Joint ALMA Observatory is operated by ESO, AUI/NRAO and
NAOJ. The National Radio Astronomy Observatory is a facility of the
National Science Foundation operated under cooperative agreement by
Associated Universities, Inc. VVG thanks support from the Chilean
Government through the Becas Chile program. KI\"O also acknowledges
funding from the Packard Foundation and an investigator award from
Simons Collaboration on the Origins of Life (SCOL). JH and RL
acknowledge support from the National Science Foundation (Grant
No. DGE-1144152).
\end{acknowledgments}

\counterwithin{figure}{section}
\appendix

\section{Channel maps}

\FigChannelMapsA{}
\FigChannelMapsB{}
\FigChannelMapsC{}
\FigChannelMapsD{}
\FigChannelMapsE{}
\FigChannelMapsF{}

\end{document}